\documentstyle [12pt] {article}

\newcommand{\gtwid}{\mathrel{\raise.3ex\hbox{$>$\kern-.75em\lower1ex
\hbox{$\sim$}}}}
\newcommand{\ltwid}{\mathrel{\raise.3ex\hbox{$<$\kern-.75em\lower1ex
\hbox{$\sim$}}}}
\newcommand{\beq}{\begin{equation}}
\newcommand{\eeq}{\end{equation}}
\newcommand{\beqs}{\begin{eqnarray}}
\newcommand{\eeqs}{\end{eqnarray}}

\catcode`@=11
\@addtoreset{equation}{section}
\@addtoreset{equation}{subsection}
\def\theequation{\ifnum\value{section}=0 \arabic{equation}\ignorespaces
\else \ifnum\value{section}=-1 A.\arabic{equation}\ignorespaces 
\else \ifnum\value{subsection}=0 \thesection.\arabic{equation}\ignorespaces
\else \thesection.\arabic{subsection}.\arabic{equation}\ignorespaces
                           \fi
                      \fi
                 \fi}
\catcode`@=12

\begin{document}

\def\thefootnote{\fnsymbol{footnote}}
\baselineskip 6.0mm

\begin{flushright}
\begin{tabular}{l}
ITP-SB-94-37    \\
hep-lat/9408020 \\
August, 1994 
\end{tabular}
\end{flushright}

\vspace{8mm}
\begin{center}
{\Large \bf Complex-Temperature Singularities of the}
\end{center}
\begin{center}
{\Large \bf Susceptibilility in the $d=2$ Ising Model \ I. Square Lattice}
\vspace{4mm}
\vspace{16mm}

\setcounter{footnote}{0}
Victor Matveev\footnote{email: vmatveev@max.physics.sunysb.edu}
\setcounter{footnote}{6}
and Robert Shrock\footnote{email: shrock@max.physics.sunysb.edu}

\vspace{6mm}
Institute for Theoretical Physics  \\
State University of New York       \\
Stony Brook, N. Y. 11794-3840  \\

\vspace{20mm}

{\bf Abstract}
\end{center}

   We investigate the complex-temperature singularities of the susceptibility 
of the 2D Ising model on a square lattice.  From an analysis of 
low-temperature series expansions, we find evidence that as one approaches 
the point $u=u_s=-1$ (where $u=e^{-4K}$) from within the complex extensions of
the FM or AFM phases, the susceptibility has a divergent singularity of the 
form $\chi \sim A_s'(1+u)^{-\gamma_s'}$ with exponent $\gamma_s'=3/2$.  The
critical amplitude $A_s'$ is calculated.  Other critical exponents are found to
be $\alpha_s'=\alpha_s=0$ and $\beta_s=1/4$, so that the scaling relation 
$\alpha_s'+2\beta_s+\gamma_s'=2$ is satisfied.  However, using exact results 
for $\beta_s$ on the square, triangular, and honeycomb lattices,
we show that universality is violated at this singularity: $\beta_s$ is
lattice-dependent.  Finally, from an analysis of spin-spin correlation
functions, we demonstrate that the correlation length and hence 
susceptibility are finite as one approaches the point $u=-1$ from within the
symmetric phase.  This is confirmed by an explicit study of high-temperature 
series expansions. 

\vspace{30mm}

\pagestyle{empty}
\newpage

\pagestyle{plain}
\pagenumbering{arabic}
\renewcommand{\thefootnote}{\arabic{footnote}}
\setcounter{footnote}{0}

\section{Introduction}
\label{intro}

    Although exact closed-form expressions for the (zero-field) free energy 
and spontaneous magnetisation of the two-dimensional Ising model were 
calculated long ago, no such expression has ever been found for the 
susceptibility, and this remains one of the classic unsolved problems in
statistical mechanics.  Any new piece of information on the susceptibility is 
thus of value, especially insofar as it specifies properties which an exact
solution must satisfy.  In particular, it is of interest to understand better
the properties of the susceptibility as an analytic function of complex 
temperature. Several years ago, some results on complex-temperature 
singularities of the susceptibility for the Ising model were reported 
\cite{ms}.  Here we continue the study of complex-temperature singularities 
of the susceptibility of the 2D Ising model. 

\section{Generalities and Discussion of Complex Extensions of Physical Phases}
\label{general}

  We consider the Ising model on a lattice $\Lambda$ at a temperature $T$ and 
external magnetic field $H$ defined in standard notation by the partition
function 
\beq
Z = \sum_{\{\sigma_i\}} e^{-\beta {\cal H}}
\label{zdef}
\eeq 
with the Hamiltonian 
\beq
{\cal H} = -J \sum_{<ij>} \sigma_i \sigma_j - H \sum_i \sigma_i 
\label{h}
\eeq
where $\sigma_i = \pm 1$ are the $Z_2$ variables on each site $i$ of the
lattice, $\beta = (k_BT)^{-1}$, $J$ is the exchange constant, 
$<ij>$ denote nearest-neighbor pairs, and the magnetic moment $\mu \equiv
1$. We shall concentrate here on the square (sq) lattice but also make some 
comments for the triangular (t) and honeycomb (hc) lattices.  We use the 
standard notation $K = \beta J$, $h = \beta H$, 
\beq
v = \tanh K
\label{v}
\eeq
\beq
z = e^{-2K} = \frac{1-v}{1+v}
\label{z}
\eeq
and 
\beq
u = z^2 = e^{-4K}
\label{u}
\eeq
It will also be useful to express certain quantities in terms of the elliptic
moduli $k_<$ and $k_> \equiv 1/k_<$.  For the square lattice these are given 
by 
\beq
k_< = \frac{1}{\sinh(2K)^{2}} = \frac{4u}{(1-u)^2}
\label{kl}
\eeq
and 
\beq
k_> = \frac{4v^2}{(1-v^2)^2}
\label{kg}
\eeq
We note the symmetries 
\beq
K \to -K \ \Rightarrow \ \{ v \to -v \ , \ \ z \to 1/z \ , \ \ u \to 1/u \ , 
\ \ k_x \to k_x \} 
\label{ksym}
\eeq
where $k_x=k_<$ or $k_>$. The reduced free energy per site is 
$f = \beta F = -N^{-1} \lim_{N \to \infty} \ln Z$ (where $N$ denotes the number
of sites on the lattice), and the zero-field susceptibility is 
$\chi_0 =\frac{\partial M(H)}{\partial H}|_{H=0}$, where $M(H)$ denotes the 
magnetisation.  Henceforth, unless otherwise stated, we
consider only the case of zero external field and drop the subscript on
$\chi_0$.  It is convenient to define the related quantity 
\beq
\bar\chi \equiv \beta^{-1}\chi
\label{barchi}
\eeq
 For the square (sq) lattice, $f(K,h=0)$ was originally 
calculated by Onsager \cite{ons}, and the expression for the spontaneous 
magnetisation $M$ was first reported by Onsager and calculated by Yang 
\cite{yang}. Solutions for $f(K,h=0)$ and $M$ were subsequently given for 
the triangular (t) and honeycomb or hexagonal (hc) lattices; for reviews, 
see Refs. \cite{domb1,domb2,mw}. We denote the critical coupling separating
the symmetric, paramagnetic (PM) high-temperature phase from the phase with
spontaneously broken $Z_2$ symmetry and ferromagnetic (FM) long-range order as
$K_c$ and recall that for the square lattice, $v_c=z_c=\sqrt{2}-1$. 

   Here we shall study the susceptibility as a function of complex (inverse) 
temperature, $K$.  For our purposes, it is important to discuss generalized 
notions of phases and thermodynamic 
quantities.  We define a complex extension of a phase as an extension, to 
complex $K$, of the physical phase which exists on a given segment of the 
real $K$ axis.  As noted, e.g., in Ref. \cite{ms}, for zero external field, 
there is an infinite periodicity in complex $K$ under certain shifts along the
imaginary $K$ axis, as a consequence of the fact that the spin-spin interaction
on each link $<ij>$ is $\sigma_i\sigma_j=\pm 1$.  In particular, there is an
infinite repetition of phases as functions of complex $K$; this infinitely
repeated set of phases is reduced to a single set by using the variables 
$v$, $z$ and/or $u$, since these latter variables have very 
simple properties under complex shifts in $K$: 
\beq
K \to K+ n i \pi \ \Rightarrow \ \{ v \to v \ , \ \ z \to z \ , \ \ u \to u \ 
,  \ \ k_x \to k_x \}
\label{shift1}
\eeq
\beq
K \to K+(2n+1)\frac{i \pi}{2} \ \Rightarrow \ \{ v \to 1/v \ , \ \ z \to -z 
\ , \ \ u \to u \ , \ \  k_x \to k_x \}
\label{shift2}
\eeq
where $n$ is an integer and, as before, $k_x=k_<$ or $k_>$.  
On a lattice with an even coordination number $q$, 
it is easily seen that these symmetries imply that the magnetisation and
susceptibility are functions only of $u$.  Because the shift (\ref{shift2})
leaves $u$ invariant while mapping $v$ to $1/v$, it maps a point in the FM
phase (and its complex extension) to itself but maps a point in the
(complex extension of the) PM phase out of this phase.  Consequently, when
studying complex-temperature properties of the model, it is more convenient to
start within the FM phase, where the various quantities of interest can be 
expressed as Taylor series in the low-temperature expansion variable $u$.
After this study, we shall proceed to investigate the properties of the 
susceptibility in the interior and boundary of the PM phase.  It is useful to
note that a given point $v_0$ or $z_0$ corresponds, in the complex $K$ plane, 
to the set
\beq
K = K_0 + n i \pi
\label{zpreimage}
\eeq
where $n \in Z$ and 
\beq
K_0 = -\frac{1}{2}\ln z_0
\label{k1}
\eeq
while a given point $u_0$ corresponds to the set 
\beq
K = K_0 + \frac{n i \pi}{2}
\label{upreimage}
\eeq
reflecting the structure of Riemann sheets of the logarithm.

     The requisite complex extensions of the physical phases can be 
seen by using the known results on the locus of points on which the 
free energy is non-analytic.  For the square lattice, these are given by the
circles \cite{fisher1}
\beq
v_\pm(\theta) = \pm 1 + 2^{1/2} e^{i \theta} \ ,  \ i.e. \ \ 
z_\pm(\omega) = \pm 1 + 2^{1/2} e^{i \omega} 
\label{circles}
\eeq
for $0 \le \theta, \ \omega < 2 \pi$. Recall that the property that this 
locus of points are circles in both the high- and low-temperature variables $v$
and $z$ follows because these variables are related by the bilinear 
conformal transformation (\ref{z}) which maps circles to circles.  For later
reference, these circles are shown in Figs. 1(a) and 1(b), respectively. 

The circles in $v$ or $z$ constitute natural boundaries, within which the free
energy is analytic but across which it cannot be analytically continued. They
thus define the complex extensions of the physical phases which occur on the
real $v$ or $z$ axes in the intervals $-v_c < v < v_c$ or 
$z_c < z < 1/z_c$ (PM); $v_c < v \le 1$ or $0 \le z < z_c$
(FM); and $-1 < v < -v_c$ or $1/z_c < z < \infty$  (AFM). 

    Using the general fact that the high-temperature expansions and 
(for discrete spin models such as the Ising model) the low-temperature 
expansions both have finite radii of
convergence, we can use standard analytic continuation arguments to establish
that not just the free energy, but also the magnetisation and susceptibility
are analytic functions within each of the complex-extended phases.  This
defines these functions as analytic functions of the respective complex
variable ($K$, $v$, $z$, $u$, or others obtained from these).  Of course,
these functions are, in general, complex away from the physical line 
$-\infty < K < \infty$.  

   We shall also need a definition of singularity exponents of a function at a
complex singular point.  In the case of real $K$, one distinguishes,
{\it a priori}, the critical exponent which describes the singular behaviour 
at a critical point approached from the symmetric, high-temperature phase from
the corresponding exponent for the approach from the broken-symmetry,
low-temperature phase.  For a singular point in the complex plane, we shall
again distinguish the critical exponents describing the singularity as 
approached from different phases. 
Thus, for the susceptibility $\bar\chi(\zeta)$, where $\zeta$
refers to one of the complex variables listed before, 
which fails to be analytic at one or more singular point(s) $\{\zeta_s\}$, if 
the leading singularity in $\bar\chi(\zeta)$ can be represented in the 
power-law form 
\beq
\bar\chi(\zeta)_{sing.} \sim |1-\zeta/\zeta_s|^{-\gamma_{s,p}}
\label{defp}
\eeq
as $\zeta$ approaches $\zeta_s$ from within the the phase $p$, we shall 
refer to $\zeta_s$ as a complex singular point and $\gamma_{s,p}$ as the 
corresponding critical or singularity exponent for the approach to $\zeta_s$
from this phase.  By analogy to standard usage for physical temperature, we
shall set $\gamma_{s,FM}=\gamma_s'$ and $\gamma_{s,PM}=\gamma_s$ to refer to 
the critical exponents at $\zeta_s$ as approached from within the complex 
extensions of the FM and PM phases, respectively.  We shall show that for the
specific point $u_s=-1$, 
$\gamma_{s,AFM}=\gamma_{s,FM}$.  Critical exponents for other quantities at
complex-temperature singular points are defined in an analogous manner. 
The locus of points $\zeta_s$ where a given function is singular in the complex
$\zeta$ plane will not, in general, be a discrete set, in contrast to the case
for the Ising model on the physical, real $K$ axis.  This is 
illustrated by the locus of points (\ref{circles}) where the free energy is
singular.  Even for a function like the magnetisation, which, in the complex FM
phase where it is nonvanishing, is an algebraic function in $\zeta=k_<$ or 
$\zeta=z$ of the form $M=\prod_{j=1}^{n_s}(\zeta-\zeta_s)^{\beta_s}$, the 
discrete points $\zeta_s$ also in general involve associated branch cuts, 
since the exponents $\beta_s$ are not integers. 

  It should be noted that a phase may exist for complex $v$ or $z$  
which is not the complex extension of any physical phase. 
An example of this phenomenon occurs in the present case; the fourth 
region, denoted O in Figs. 1(a) and 1(b) constitutes such a phase.  

    In contrast to the usual ferromagnetic critical point the Ising model, 
which can be approached only 
within the PM phase or FM phase (and similarly, the AFM critical point of this
model, which can be approached only from within the PM or AFM phase), a general
complex singularity may be approached from within more than two phases.  For
example, in Fig. 1, the singularities at $v=\pm i$, or equivalently, 
$z=\pm i$, can be approached from within the PM, FM, or AFM complex-extended
phases, or, indeed, from the region O which is not analytically connected to
any physical phase.

    Since for the square lattice $\bar\chi(z)$ has the symmetry noted above,
$\bar\chi(z)=\bar\chi(-z)$, it is useful to display the 
complex-extended phases as functions of $u$.  Under the conformal 
transformation $u=z^2$, the circles in Fig. 1(b) are mapped to a single 
curve, which is a type of lima\c{c}on of Pascal, defined by 
\beq
Re(u) = 1 + 2^{3/2}\cos \omega + 2 \cos 2\omega
\label{reu}
\eeq
\beq
Im(u) = 2^{3/2}\sin \omega + 2 \sin 2\omega
\label{limacon}
\eeq
traced out completely for $0 \le \omega < 2\pi$.  In this variable, there are
three complex-extended phases, as shown in Fig. 1(c): PM, FM, and AFM. The
mapping from $z$ to $u$ reduced the number 
of complex-extended phases from the four which are present in the variable $z$
or $v$ to three; since the points in the O phase are related to those in the
complex-extended PM phase by $z \to -z$, these two phases are mapped to a
single phase in the $u$ plane. The points $z = \pm i$, at which the PM, FM, 
AFM, and O phases are all contiguous, are mapped to the single point
$u=u_s=-1$ in the $u$ plane.  In the terminology of algebraic geometry, the 
point $u_s=-1$ is a singular point, specifically a multiple point of index 2, 
of the lima\c{c}on 
(\ref{limacon}) forming the natural boundary between the complex phases, 
whereas all other points on this curve, including the PM-FM critical point 
$u_c$ and the PM-AFM critical point at $u=1/u_c$, are regular
(ordinary) points of the curve. Here, a multiple point of index $n$ of a 
curve $C$ is a point through which $n$ arcs of $C$ pass (see, e.g., Ref. 
\cite{alg}).  For later reference, the physical critical points are
$u_c=3-2\sqrt{2}=0.171572875...$ separating the PM and FM phases and 
$u_c^{-1}=3+2\sqrt{2}=5.8284271...$ separating the PM and AFM phases.  In
the complex $K$ plane, these correspond to the infinite set of critical points
$K_c=\pm (1/2)\ln(1+\sqrt{2}) + n i \pi/2$, where $n \in Z$.  Under the 
transformation $u \to 1/u$, the complex-extended PM phase maps onto itself, 
while the complex-extended FM phase maps to the AFM phase, and vice versa. 

   Finally, in terms of the elliptic moduli, the natural boundaries have the
very simple form of the unit circle in the complex $k_<$ or $k_>$ planes:
\beq
k_< = 1/k_> = e^{i\theta}
\label{kcircle}
\eeq
with $0 \le \theta < 2\pi$.  These incorporate the symmetries 
\beq
u \to 1/u \ \Rightarrow \ k_x \to k_x
\label{uinversion}
\eeq
\beq
v \to 1/v \ \Rightarrow \ k_x \to k_x
\label{vinversion}
\eeq
where $k_x$ denotes $k_<$, $k_>$, or $\kappa$.  Given the inversion symmetry 
(\ref{uinversion}), it follows that the transformation (\ref{kl}) 
from $u$ to $k_<$ maps both the complex-extended FM and AFM phases onto the 
same region, which is the interior of the unit circle in the complex $k_<$ 
plane.  The complex-extended PM phase is mapped to the exterior of this circle.
Under the mapping, the actual lima\c{c}on in the $u$ plane wraps around the 
unit circle in the $k_<$ plane twice.  In particular, both the PM-FM critical 
point $u_{c,sq}$ and the PM-AFM critical point $1/u_c$ are mapped to 
the single point, $k_<=k_>=1$.  The 
complex-temperature singular point $u_s=-1$ is mapped to $k_<=k_>=-1$. 

   Having discussed these preliminaries, we proceed to study the
susceptibility. 

\section{Analysis of Low-Temperature Series}
\label{analysis}

\subsection{Analysis of Series for $\bar\chi_r$ in the Variable $u$}
\label{useriesanalysis}

     The low-temperature series expansion for $\bar\chi$ for the Ising model 
on the square lattice is 
\beq
\bar\chi = 4u^2 \Bigl (1 + \sum_{n=1}^{\infty} c_n u^n \Bigr )
\label{chiuseries}
\eeq
This expansion has a finite radius of convergence and, by analytic 
continuation from the physical low-temperature interval $0 \le u < u_c$, 
applies throughout the complex extension of the FM phase.  Since the 
factor $4u^2$ is known exactly, it is convenient to study the reduced (r) 
function 
\beq
\bar\chi_r \equiv 2^{-2}u^{-2}\bar\chi = 1 + \sum_{n=1}^{\infty}c_n u^n  
\label{chir}
\eeq
The expansion coefficients $c_n$ were calculated to order $n=9$ in 1971 by 
the King's College group \cite{kc1} and were extended to order $n=21$ by 
Baxter and Enting in 1978 \cite{be} (with exact coefficients up to $n=19$ and
nearly exact $n=20$ and 21 terms). Very recently, the $c_n$'s have been 
calculated to order $n=26$ (i.e. $\bar\chi$ to $O(u^{28})$) by 
Briggs, Enting, and Guttmann \cite{beg}, as part of a general calculation 
of low-temperature series for $q$-state Potts 
models with $q=2$ to $q=10$ on the square lattice. We have carried out a dlog 
Pad\'{e} analysis of this series to investigate the singular behaviour of the
susceptibility in the complex $u$ plane. (For reviews of this method, see 
Ref. \cite{pade}.)  As one approaches a complex singular point denoted 
$s$ on the boundary of the complex-extended FM phase from within this phase, 
$\bar\chi$ is assumed to have the leading singularity ($s$)  
\beq
\bar\chi(u) \sim A_s'|1-u/u_s|^{-\gamma_s'}\biggl ( 1+a_{1,s}|1-u/u_s| + ...
\biggr )
\label{singform}
\eeq
where $A_s'$ and $\gamma_s'$ denote, respectively, the critical amplitude and
the corresponding critical exponent, and the dots $...$ 
represent analytic confluent
corrections.  One may observe that we have not included non-analytic confluent
corrections to the scaling form in eq. (\ref{singform}).  The reason is that, 
although such terms are generally present at critical points in statistical 
mechanical models, previous studies have indicated that they are very 
weak or absent for the usual critical point of the 2D Ising model 
\cite{conflu1,conflu2}.  The dlog Pad\'{e} study then directly gives $u_s$ 
and $\gamma_s'$.  As noted above, the prefactor $4u^2$ is known and is 
analytic, so we actually carry out the dlog Pad\'{e} study on $\bar\chi_r$.  
This study yields evidence for a divergent branch point singularity at a
particular complex-temperature point, which we denote $u_s$.  (The use of $u_s$
above referred to generic complex-temperature point(s), of which there might,
{\it a priori}, be more than one; henceforth, we use this symbol to refer to
the specific point found from the Pad\'{e} study.) 
The results for $u_s$ and $\gamma_s'$ from the diagonal and near-diagonal 
approximants are listed in Tables 1 and 2, starting with series for 
$\bar\chi_r$ to $O(u^{12}$) and going up to $O(u^{26})$.
\begin{table}
\begin{center}
\begin{tabular}{|c|c|c|c|c|c|} \hline \hline  & & & & & \\
$N$ & $[(N-2)/N]$ & $[(N-1)/N]$ & $[N/N]$ & $[(N+1)/N]$ & $[(N+2)/N]$ \\
 & & & & & \\
\hline \hline 
6 & $-$ & $-1.05026$ & $-1.07608^{*}$ & $-0.992720$ & $-0.990763$ \\ \hline
7 & $-1.09018^*$ & $-1.03635$ & $-0.990673$  & $-0.992677^*$ & $-1.00752^*$ \\
\hline
8 & $-1.00060$ & $-1.02216$ & $-1.00363^*$ & $-0.997028$ & $-0.988543$ \\ 
\hline
9 & $-1.00762^*$ & $-1.18678$ & $-0.981169$ & $-0.9954185^*$ & $-0.9963225^*$ 
\\ \hline
10 & $-0.983770$ & $-1.13876^*$ & $-0.996264^*$ & $-0.995312^*$ & 
$-0.998649^*$ \\ \hline
11 & $-0.999211$ & $-1.00398$ & $-0.998086^*$ & $-0.9977515^*$ & 
$-0.998400^*$ \\ \hline
12 & $-0.999976^*$ & $-0.997679^*$ & $-0.998044^*$ & $-0.997225^*$ & $-$ \\ 
\hline
13 & $-0.999073^*$ & $-0.996210^*$ &  $-$     &   $-$     &   $-$        \\ 
\hline \hline
\end{tabular}
\end{center}
\caption{Values of $u_s$ from Pad\'{e} approximants to low-temperature series
for $\bar\chi_r$ starting with the series to $O(u^{12})$. Superscript $*$
indicates that the approximant has one or more nearly coincident 
pole-zero pair(s) closer to the origin than $u_s$. Our criterion for near
coincidence is that $|u_{pole}-u_{zero}| < 10^{-4}$.}
\label{table1}
\begin{center}
\begin{tabular}{|c|c|c|c|c|c|} \hline \hline  & & & & & \\
$N$ & $[(N-2)/N]$ & $[(N-1)/N]$ & $[N/N]$ & $[(N+1)/N]$ & $[(N+2)/N]$ \\
 & & & & & \\
\hline\hline 
6  & $-$      & 1.975    & 2.174$^*$ & 1.474     & 1.455     \\ \hline
7  & 2.297$*$ & 1.879    & 1.454     & 1.474$^*$ & 1.631$^*$ \\ \hline
8  & 1.563    & 1.771    & 1.583$^*$ & 1.513     & 1.419     \\ \hline
9  & 1.628$^*$ & 1.533   & 1.321   & 1.497$^*$  & 1.508$^*$  \\ \hline
10 & 1.358 & 1.732$^*$ & 1.507$^*$ & 1.496$^*$ & 1.536$^*$   \\ \hline
11 & 1.545 & 1.605 & 1.528$^*$ & 1.524$^*$ & 1.533$^*$       \\ \hline
12 & 1.554$^*$ & 1.523$^*$ & 1.528$^*$ & 1.517$^*$  & $-$    \\ \hline
13 & 1.543$^*$ & 1.503$^*$ &  $-$       &  $-$    &   $-$    \\ 
\hline  \hline
\end{tabular}
\end{center}
\caption{Values of $\gamma_s'$ from Pad\'{e} approximants to low-temperature
series for $\bar\chi_r$ starting with the series to $O(u^{12})$.}
\label{table2}
\end{table}
We do not find evidence for any other complex-temperature singularities 
within the range described by the small-$u$ expansion.  
 From this Pad\'{e} analysis of the low-temperature series, we infer the 
values 
\beq
u_{s} = -0.998 \pm 0.002
\label{us}
\eeq
\beq
\gamma_s' = 1.52 \pm 0.06
\label{gammas}
\eeq
where the uncertainties are estimates.  These results suggest the
conclusion that as $u$ approaches the point 
\beq
u_s=-1
\label{usconc}
\eeq
from within the FM phase, $\bar\chi$ has a divergent singularity with exponent
\beq
\gamma_s' = \frac{3}{2}
\label{gammasconc}
\eeq
As noted above, this point $u_s=-1$ corresponds to the two points $z_s = -v_s =
\pm i$; in the complex $K$ plane it corresponds to the infinite set of points 
given by 
\beq
K_s = -\frac{i \pi}{4} + \frac{n i \pi}{2} 
\label{ks}
\eeq
with $n \in Z$.

   Below we shall show that the critical exponent for  the inverse correlation
length (mass gap) describing row or column connected spin-spin correlation 
functions at this singular point is $\nu_{s,row}'=1$.  A naive 
complex-temperature analogue of the usual argument for the scaling
relation $\nu'(2-\eta)=\gamma'$, in conjunction with our inferred value
of $\gamma_s'$ in eq. (\ref{gammasconc}), would lead to the further inference 
that the exponent describing the asymptotic decay of the row or column 
connected 2-spin correlation function at the singular point $u_s=-1$ is 
$\eta_{s,row}'=1/2$ (where we
append the prime to indicate that the calculation of the spin-spin correlation
function involves a limit from the complex FM phase).  However, we shall show
that the situation near the complex-temperature singular point $u=-1$ 
is considerably more complicated than the case at the
physical critical point with its simple scaling relations 
$\nu'(2-\eta)=\gamma'$ and $\nu(2-\eta)=\gamma$.  Among other things, we shall
show that the correlation length and $\bar\chi$ are finite when one 
approaches $u=-1$ from the complex PM phase. 

   The dlog Pad\'{e} study did not yield any evidence for other complex 
singularities; i.e., it did not give poles whose positions were highly stable 
as one varied the orders of the approximants.  As usual, the values of the 
position of the singular point vary less among the Pad\'{e} entries than the 
values of the exponent.  Also, as expected, the values of $\gamma_s'$ show 
less scatter 
in the higher-order Pad\'{e} entries than in the lower order entries.  It is
true, however, that these values of $\gamma_s'$ do exhibit more scatter than 
the values of the usual susceptibility exponent $\gamma'$ for the
PM-FM critical point $u_c$.  To make this comparison quantitative, it is
sufficient to show the values of $u_c$ and $\gamma'$ extracted from just
the diagonal Pad\'{e} entries; these are given in Table 3. 
\begin{table}
\begin{center}
\begin{tabular}{|c|c|c|c|} \hline \hline  & & & \\
$[N/N]$ & $u_{c,ser.}$ & $|u_{c,ser.}-u_c|/u_c$ & $\gamma'$ \\
 & & & \\
\hline \hline 
[6/6]      & 0.17154017 & $1.9 \times 10^{-4}$    & 1.740  \\ \hline
[7/7]      & 0.17156038  & $0.73 \times 10^{-4}$   & 1.745  \\ \hline
[8/8]  & 0.17146527$^*$  & $6.3 \times 10^{-4}$    & 1.791$^*$ \\ \hline  
[9/9]  & 0.17156858$^*$  & $2.5 \times 10^{-5}$    & 1.747$^*$ \\ \hline
[10/10]    & 0.17157013  & $1.6 \times 10^{-5}$    & 1.748  \\ \hline
[11/11]    & 0.17157232  & $3.2 \times 10^{-6}$    & 1.749  \\ \hline
[12/12] & 0.17157423$^*$  & $7.9 \times 10^{-6}$   & 1.751$^*$  \\ \hline
exact  & 0.171572875..   &        0                & 1.750  \\
\hline \hline
\end{tabular}
\end{center}
\caption{Values of $u_{c,ser.}$ and $\gamma'$ from diagonal Pad\'{e} 
approximants to low-temperature series for $\bar\chi_r$ starting with the
series to $O(u^{13})$.}
\label{table3}
\end{table}

\subsection{Analysis of Series for $g(u)$ }

   Clearly, a property of Pad\'{e} approximants which is crucial for our study
is their sensitivity to singularities which are not the closest to the origin
of the Taylor series expansion.  In order to explore the possibility of 
obtaining a more sensitive probe of the complex singular point,
we have also carried out a similar study of a series with the physical 
singularity removed.  In order to keep the coefficients rational, we actually
multiply $\bar\chi$ by the factor $[(u-u_c)(u-1/u_c)]^{7/4} =
(1-6u+u^2)^{7/4}$ and thus study
\beq
g(u) \equiv (1-6u+u^2)^{7/4}\bar\chi_r(u)
\label{g}
\eeq
This is an old technique (see, e.g., Ref. \cite{domb2}).  Note that the
spurious finite branch-point singularity introduced at the PM-AFM critical
point has no effect on our analysis, since the small-$|u|$ series for
$\bar\chi$ and $g$ only apply in the FM region, which is not contiguous with
the region of the PM-AFM critical point in the $u$ plane.  The results of our
Pad\'{e} analysis of the series for $g(u)$ are given in Tables 4 and 5. 
\begin{table}
\begin{center}
\begin{tabular}{|c|c|c|c|c|c|} \hline \hline  & & & & & \\
$N$ & $[(N-2)/N]$ & $[(N-1)/N]$ & $[N/N]$ & $[(N+1)/N]$ & $[(N+2)/N]$ \\
 & & & & & \\
\hline \hline 
6 & $-$ & $-1.05279$ & $-1.07822^*$ & $-0.991486$ & $-1.001045$  \\ \hline
7 & $-1.09002^*$ & $-1.04502$ & $-0.999217$ & $-0.989684^*$ & $-1.00833^*$ \\ 
\hline
8 & $-1.010665$ & $-1.02141$ & $-1.00384^*$ & $-0.996939$ & $-0.989857$ \\ 
\hline
9 & $-1.01188^*$ & $-1.34302$ & $-0.985198$ & $-0.995437$ & $-0.996976$ \\ 
\hline
10 & $-0.988462$ & $-1.26478^*$ & $-0.996803$ & $-0.995048^*$ & $-0.998644^*$ 
\\ \hline
11 & $-1.00156$ & $-1.00354$ & $-0.998084$ & $-0.997731$ & $-0.998407^*$ \\ 
\hline
12 & $-1.001685^*$ & $-0.997653$ & $-0.998042^*$ & $-0.996954^*$ & $-$ \\ 
\hline
13 & $-0.999656^*$ & $-0.994374^*$ &  $-$     &   $-$     &   $-$   \\ 
\hline \hline
\end{tabular}
\end{center}
\caption{Values of $u_s$ from Pad\'{e} approximants to 
$g(u)=(1-6u+u^2)^{7/4}\bar\chi_r$, starting with the series for $g(u)$ to 
order $O(u^{12})$.}
\label{table4}
\end{table}
\begin{table}
\begin{center}
\begin{tabular}{|c|c|c|c|c|c|} \hline \hline  & & & & & \\
$N$ & $[(N-2)/N]$ & $[(N-1)/N]$ & $[N/N]$ & $[(N+1)/N]$ & $[(N+2)/N]$ \\
 & & & & & \\
\hline\hline 
6  & $-$  & 1.998  & 2.179$^*$ & 1.464   & 1.561     \\ \hline
7  & 2.278$^*$ & 1.950    & 1.540   & 1.449$^*$ & 1.640$^*$ \\ \hline
8  & 1.664    & 1.769    & 1.585$^*$ & 1.512     & 1.434    \\ \hline
9  & 1.674$^*$ & 1.356   & 1.372   & 1.497  & 1.516  \\ \hline
10 & 1.418 & 1.490$^*$ & 1.513 & 1.493$^*$ & 1.536$^*$   \\ \hline
11 & 1.575 & 1.600 & 1.528 & 1.524 & 1.533$^*$       \\ \hline
12 & 1.576$^*$ & 1.523 & 1.528$^*$ & 1.514$^*$  & $-$    \\ \hline
13 & 1.551$^*$ & 1.481$^*$ &  $-$       &  $-$    &   $-$    \\ 
\hline \hline
\end{tabular}
\end{center}
\caption{Values of $\gamma_s'$ from Pad\'{e} approximants to
$g(u)=(1-6u+u^2)^{7/4}\bar\chi_r$, starting with the series for $g(u)$ to 
order $O(u^{12})$.} 
\label{table5}
\end{table}
As one can see from these tables, the resultant values for $u_s$ and
$\gamma_s'$ are in very good agreement with those from our analysis of the
series for $\bar\chi_r$.  One improvement that occurs is a slight reduction of
the number of entries with nearly coincident pole-zero pairs, which should
increase the accuracy of the results somewhat. 

\subsection{Analysis of Series for $\bar\chi$ in the Variable $k_<$}

    Finally, it is useful to transform the series for the susceptibility from
the usual low-temperature variable $u$ to the elliptic modulus $k_<$ and to
study the resultant series.  An important motivation for this is that
$\bar\chi$ is given formally by a sum over all connected correlation functions,
and these correlation functions, which can be computed exactly in terms of
certain Toeplitz determinants \cite{kauff,mpw}, have explicit forms which are
polynomials in the complete elliptic integrals $K(k_x)$ and $E(k_x)$ \cite{gs},
where $k_x = k_<$ in the FM and AFM phases and $k_x=k_>$ in the PM phase. 
The variables $k_<$ and $k_>$ are thus natural ones for low- and 
high-temperature series expansions of $\bar\chi$, respectively.  We therefore 
have transformed the known small-$|u|$ series to one in $k_<$, which takes the
form 
\beq
\bar\chi = 2^{-2}k_<^2 \Bigl ( 1 + \sum_{n=1}^{\infty}
c_n' k_<^n \Bigr )
\label{klseries}
\eeq
The series in parentheses defines a reduced function
$\bar\chi_r=4k_<^{-2}\bar\chi$ as before. 
Since as a function of $u$, $k_<$ has the expansion near $u=u_s=-1$ 
\beq
k_< = -1 + \frac{1}{4}(u+1)^2 + O((u+1)^3)
\label{klexpansion}
\eeq
with no linear term, it follows that in the variable $k_<$, the singular form 
of $\bar\chi$ corresponding to eq. (\ref{singform}), as $k_<$ approaches the
point $k_<=-1$ from within the FM or AFM phase (i.e. from within the interior
of the unit circle in the complex $k_<$ plane) is 
\beq
\bar\chi(k_<) \sim B_s'|1+k_<|^{-\gamma_s'/2}\biggl ( 1+b_1|1+k_<| + ...
\biggr )
\label{singformk}
\eeq
where $B_s'$ is the critical amplitude for this expression of the singularity 
in terms of the variable $k_<$.  We have performed a dlog Pad\'{e} analysis of
the series in $k_<$ for $\bar\chi_r$, i.e., an analysis of the function 
$d\ln \bar\chi_r/dk_<$.  Our results for the diagonal entries are given in 
Table 6. 
\begin{table}
\begin{center}
\begin{tabular}{|c|c|c|c|} \hline \hline  & & & \\
$[N/N]$ & $(k_<)_{s,ser.}$ & $|(k_<)_{s,ser.}+1|$ & $\gamma_s'$ \\
 & & & \\
\hline \hline 
[6/6]   & $-0.998766$     & $ 1.2 \times 10^{-3}$   & 1.513  \\ \hline
[7/7]   & $-0.999940$     & $0.60 \times 10^{-4}$   & 1.552  \\ \hline
[8/8]   & $-1.000012$     & $1.2 \times 10^{-5}$    & 1.555  \\ \hline  
[9/9]   & $-0.999734$     & $2.7 \times 10^{-4}$    & 1.543  \\ \hline
[10/10] & $-0.999819^*$   & $1.8 \times 10^{-4}$    & $1.546^*$  \\ \hline
[11/11] & $-0.999873$     & $1.3 \times 10^{-4}$    & 1.549  \\ \hline
[12/12] & c.p.            &  $-$                    &  $-$  \\ 
\hline \hline
\end{tabular}
\end{center}
\caption{Values of $(k_<)_{s,ser.}$ and $\gamma_s'$ from diagonal Pad\'{e} 
approximants to low-temperature series for $\bar\chi_r$ starting with the
series to $O(k_<^{13})$. c.p. denotes a complex pair of poles close to
$-1$.  Superscript $*$ indicates that the approximant has one or more nearly 
coincident pole-zero pair(s) closer to the origin than $(k_<)_s=-1$. As before,
our criterion for near-coincidence is that 
$|(k_<)_{pole}-(k_<)_{zero}| < 10^{-4}$. }
\label{table6}
\end{table}
The results from the series in $k_<$ agree very well with 
those which we obtained from the other series.  The analysis of the series in
$k_<$ also gives results for the regular PM-FM critical point at $k_<=1$ and
the associated exponent $\gamma'$ of comparable accuracy to that of the series
in $u$. 
We have also carried out an analysis of the series for $\bar\chi_r$ in the
variable $u$ using differential
approximants (for further details on this method, see section 7 below).  This
yields results close to those in Table 6.

   One important new piece of information can be obtained from our analysis 
of the series for $\bar\chi$ in the variable $k_<$ near the singular point 
$k_<=(k_<)_s=u_s=-1$: this is an answer to the question of whether the critical
exponent is the same when one approaches this point from within the interior of
the complex-extended FM phase and from within the complex-extended AFM
phase. Recall from Fig. 1(c) that in the $u$ plane, these approaches are
distinct, since the complex FM and AFM phases lie on opposite sides of $u_s$. 
Since the complex FM phase is the one which contains the origin in the $u$ 
plane, our analysis 
of the small-$|u|$ series could only determine the singular behaviour as $u$ 
approached $u_s$ from within the complex FM phase.  However, since the complex
 FM and AFM phases are mapped onto each other in the $k_<$ plane, our analysis
of the series in this variable shows that the exponent $\gamma_s'$ is the same
for the approach to $u_s=(k_<)_s=-1$ from within both the FM and AFM phases (as
has been implicit in our notation). 

    Since the low-temperature series for $\bar\chi$ is not usually given in 
terms of the variable $k_<$, we note 
that it has an interesting feature.  The first few terms are 
\beq
\bar\chi_r = 1 + k_< + \frac{13}{2^3}k_<^2 + \frac{13}{2^3}k_<^3 + 
\frac{139}{2^6}k_<^4 + \frac{139}{2^6}k_<^5 + \frac{685}{2^8}k_<^6 
+ \frac{2739}{2^9}k_<^7 + \frac{51603}{2^{14}}k_<^8 + O(k_<^9)
\label{klterms}
\eeq
Near the point $u_s=(k_<)_s=-1$, the terms up to order $O(k_<^5)$ of the 
series exactly 
cancel among each other in a successive pairwise fashion, so that the first
nonzero terms in the series for $\bar\chi_r$ start at order $O(k_<^6)$ 
(i.e. for $\bar\chi$ at $O(k_<^8)$).  To say it differently, these first six
terms can be expressed as $(1+k_<)(1+(13/2^3)k_<^2+(139/2^6)k_<^4)$. 
One can interpret this as being a hint of a structure
which persists to all orders in the exact susceptibility; with this motivation,
one can add and subtract terms in higher orders so as to put $\bar\chi_r$ 
in the form 
\beq
\bar\chi = (1+k_<)f_1 - 2^{-12}k_<^7 f_2
\label{chicon}
\eeq
Computing the functions $f(k_<)_j$, $j=1,2$ and defining the convenient
variable 
\beq
y \equiv 2^{-4}k_<^2
\label{ydef}
\eeq
we find
\beqs
f_1 = & 1 + 26y + 556y^2 + 10960y^3 + 206412y^4 + 3775480y^5 + 67668304y^6 + 
1194824896y^7 \nonumber \\ 
      & + 20856575980y^8 + 360778731928y^9 + 6195017443856y^{10} \nonumber \\
      & + 105730294168640y^{11} + 1795278082108368y^{12}+O(y^{13})
\label{f1}
\eeqs
and 
\beqs
f_2 = & 1 + 33y + 770y^2 + 15650y^3 + 296006y^4 + 5363335y^5 + 94504364y^6 +
1633461856y^7 \nonumber \\
      & + 27844153964y^8+469735545258y^9+O(y^{10}) 
\label{f2}
\eeqs
We have performed a Pad\'{e} analysis of the functions $d \ln f_j/dy$.  We find
strong evidence of a singularity in $f_1$ of the form 
$f_1 \sim |1-k_<^2|^{-7/4}$ as $k_<^2$ increases toward 1 from below. Combining
this with the $(1-k_<)$ prefactor, it follows that the singularity in
$\bar\chi$ arising from the $f_1$ term can be written as 
\beq
(1+k_<)f_1 \sim |1-k_<|^{-7/4}|1+k_<|^{-3/4}
\label{f1sing}
\eeq
as $k_<^2$ approaches 1 from below. For example, the $[6/6]$ Pad\'{e}
approximant gives $16y=k_<^2=0.999844$ as the position of the singularity, and
$1.745$ as the exponent. The results from the study of $f_2$ are also 
consistent with this conclusion for the singularity in $\bar\chi$.  Recalling
that $|1+k_<|^{-3/4} = const \times |1+u|^{-3/2}$ as $k_< \to -1$ or
equivalently, as $u \to -1$,
one sees that the singular form (\ref{f1sing}) agrees very nicely with 
our determination of the complex-temperature singularity by analyses of the
small-$|u|$ series for $\bar\chi_r(u)$, $g(u)$, and $\bar\chi_r(k_<)$ given 
above.

\subsection{Critical Amplitude at $u_s$}

   In order to calculate the critical amplitude $A_s'$ in the susceptibility as
one approaches $u=-1$ from within the FM phase, we compute the series for 
$(\bar\chi_r)^{1/\gamma_s'}$.  Since the exact function
$(\bar\chi_r)^{1/\gamma_s'}$ has a simple pole at $u_s$, one performs the
Pad\'{e} analysis on the series itself instead of its logarithmic
derivative. The residue at this pole is $-u_s(A_{r,s}')^{1/\gamma_s'}$, 
where $A_{s,r}'$ denotes the critical amplitude for $\bar\chi_r$.  Using our
inferred value $\gamma_s=3/2$ to calculate the series and the value $u_s=-1$ to
extract $A_{r,s}'$, we finally multiply by the prefactor to obtain
$A_s'=4u_s^2A_{r,s}'=-4A_s'$.  (Alternatively, one could extract
$(A_{r,s}')^{1/\gamma_s'}$ from the residue by dividing by the measured pole
position, $u_{s,ser.}$ from the given Pad\'{e} rather than the inferred 
exact position, and could use the prefactor $4u_{s,ser}^2$ to get $A_s'$; the
differences between the two methods are quite small and vanish 
asymptotically; these differences are incorporated in the final uncertainty
which is quoted for the critical amplitude.)  Our results from the diagonal
Pad\'{e} entries are listed in Table 7. 
\begin{table}
\begin{center}
\begin{tabular}{|c|c|c|} \hline \hline  & & \\
$[N/N]$ & $u_s$  & $R_s=-u_s(A_{r,s}')^{2/3}$ \\
 & & \\
\hline \hline 
[6/6]   & $-0.993282$    & 0.128915 \\ \hline
[7/7]   & $-0.9942585$   & 0.129496 \\ \hline
[8/8]   & $-0.994204$    & 0.129460 \\ \hline
[9/9]   & $-0.994088$    & 0.1293865 \\ \hline
[10/10] & $-0.995080$    & 0.130178  \\ \hline
[11/11] & $-0.993497$    & 0.129294  \\ \hline
[12/12] & c.p.           & c.p.     \\ \hline
\hline \hline
\end{tabular}
\end{center}
\caption{Values of $(A_s')^{2/3}$ from Pad\'{e} approximants to 
small-$|u|$ series for $(\bar\chi_r)^{1/\gamma}$. c.p. indicates a complex
pair of poles near to $u_s=-1$. } 
\label{table7}
\end{table}
 From this analysis, we calculate
\beq
A_s' = 0.186 \pm 0.001
\label{as}
\eeq
where the quoted uncertainty is an estimate.  

    This value may be compared with
the low-temperature critical amplitude in this model at the usual PM-FM
critical point, $u_c$, defined by $\bar\chi(u) \sim A_c'|1-u/u_c|^{-7/4}$ as 
$u \to u_c$ from below. $A_c'$ was determined first by analysis of 
low-temperature series expansions \cite{domb2} and subsequently to higher 
accuracy by analytic methods \cite{conflu2} 
\footnote{Ref. \cite{conflu2} actually gives the critical amplitude $A_{c,T}'$
defined by $\bar\chi(T) \sim A_{c,T}'|1-T_c/T|^{-7/4}$; we have converted this
to $A_c'$ here.} to be $A_c' = 0.068865538..$.  Using our determination of
$A_s'$, it follows that $A_s'/A_c' = 2.701 \pm 0.015$.  We note that this is 
consistent, to within the numerical accuracy, with the analytic relation 
$A_s'/A_c' = (-\ln u_c)^{7/4}=2.6966995..$. 

\section{Singular Behaviour of Other Quantities at $u_s=-1$}
\label{others}

\subsection{Specific Heat}
\label{csection}

   In this section we extract the singular behaviour of the exactly known
thermodynamic quantities at the complex-temperature singular point $u_s=-1$. We
begin with the specific heat.  It is convenient to consider $K^{-2}C$, which 
(in units with $k_B \equiv 1$) is given by \cite{ons} 
\beq
K^{-2}C = \frac{4}{\pi}\biggl ( \frac{1-\kappa'}{\kappa^2} \biggr )
\biggl [ 2 \Bigl (K(\kappa)-
E(\kappa) \Bigr ) - (1-\kappa') \Bigl (\frac{\pi}{2}+\kappa'K(\kappa) 
\Bigr ) \biggr ]
\label{ceq}
\eeq
where the elliptic modulus is 
\beq
\kappa= \frac{2}{(k_>^{1/2}+k_>^{-1/2})} = \frac{2}{(k_<^{1/2}+k_<^{-1/2})} =  
 \frac{4u^{1/2}(1-u)}{(1+u)^2}
\label{kappa}
\eeq
and its complementary modulus is $\kappa'=(1-\kappa^2)^{1/2}$.  We use the 
standard convention that the branch cut for the complete elliptic integrals 
runs from $m \equiv \kappa^2 =1$ to $m=\infty$ along the positive real axis in
the complex $m$-plane. From eq. (\ref{kappa}), it follows that as 
$u$ approaches $-1$ with $Im(u)$ positive,  $\kappa \to i\infty$ (and if
$Im(u) < 0$, then $\kappa \to -i \infty$, taking the usual convention for the
branch cut for the square root), and thus $\kappa' \to \infty$. 
 From inspection of (\ref{ceq}), it is clear
that as $u \to -1$, $C$ diverges, with the leading divergence arising from the
last term, $-(4/\pi)(\kappa'^3/\kappa^2)K(\kappa) \to (4/\pi)\kappa'
K(\kappa)$. Using the identity (see e.g., Ref. \cite{grad})
$\kappa' K(\kappa) = K(i\kappa/\kappa')$ and the fact that as $\lambda \to 1$, 
$K(\lambda) \to (1/2)\ln(16/(1-\lambda^2))$, we can express the most singular
term as $(4/\pi)\ln(4|\kappa|)$ as $u \to -1$.  Next, using the fact that 
near $u=-1$, $1/\kappa=-(i/8)(1+u)^2 + O((1+u)^3$, 
we find, finally, that the leading divergence in $C$ as $u \to -1$ is
\beq
C \sim \frac{8}{\pi}K_s^2\ln \Bigl ( \frac{1}{|1+u|} \Bigr )
\label{csing}
\eeq
Taking the value of $K_s$ on the first Reimann sheet in eq. (\ref{ks}), i.e.,
$K_s=-i\pi/4$, this becomes 
\beq
C \sim -\frac{\pi}{2} \ln \Bigl ( \frac{1}{|1+u|} \Bigr )
\label{csing2}
\eeq
Since the elliptic integrals and also the factor $\kappa'$ only depend on 
$\kappa^2$, the leading singularity is the same whether $u$ approaches $u_s=-1$
with $Im(u)$ positive, negative, or zero, and also whether the approach occurs
from within the complex FM, AFM, or PM phases.  The logarithmic divergence in
$C$ at $u_s$ is evidently of the same type as the divergence at the physical
PM-FM and PM-AFM critical points.  Note, however, that at these latter 
points, $\kappa \to 1$ ($\kappa' \to 0$) so that the 
$\kappa'K(\kappa)$ term (which gives the leading divergence at $u=-1$) 
vanishes, and the divergence arises instead from the first term in the square
brackets of eq. (\ref{ceq}), $2K(\kappa)$. Another obvious difference is that,
while the 
specific heat is required to be positive at physical temperatures, it is, in
general, complex at complex-temperature points, and the critical amplitude 
at $u_s$ is real but negative. 
The critical exponents corresponding to this logarithmic divergence in $C$ 
at $u_s$ are
\beq
\alpha_s = \alpha_s' = 0 
\label{alphas}
\eeq

\subsection{Magnetisation}
\label{magsection}

    Next, we make use of the exactly known expressions for the spontaneous
magnetisation $M$ to analyse the behaviour as a function of complex 
temperature. In particular, we shall extract the critical exponent $\beta_s$ 
at the complex-temperature singular point $u=u_s=-1$. 

 For the square lattice, for 
real temperature, $M$ vanishes for $K < K_{c,sq}$ (where for clarity we restore
here the subscript indicating the lattice type) and, for $K_{c,sq} < K < 
\infty$ is given by \cite{yang} 
$M=M_{sq}=(1-k_<^2)^{1/8}$, or, in terms of $u$, 
\beq
M_{sq} = \frac{(1+u)^{1/4}(1-6u+u^2)^{1/8}}{(1-u)^{1/2}}
\label{msq}
\eeq
What is normally discussed is the vanishing of $M$ at the usual PM-FM
critical point, $u_c=3-2\sqrt{2}$, with exponent $\beta=1/8$.  However, as we
discussed in section \ref{general}, the function describing the magnetisation
for positive temperature can be analytically continued throughout the complex
extension of the FM phase, up to the boundaries of this phase, which, for the
square lattice are specified by the lima\c{c}on (\ref{limacon}).  
Carrying out this analytic continuation, one sees two 
important results: (1) the only point, other than the physical PM-FM critical
point, where $M$ vanishes continuously, is at $u=u_s=-1$.  Defining an 
associated critical exponent as 
\beq
|M| \sim const. \times |1-u/u_s|^{\beta_s}
\label{betadef}
\eeq
as $u$ approaches $u_s$ from within the complex-extended FM phase, we find the
value 
\beq
\beta_{s,sq} = \frac{1}{4}
\label{betas}
\eeq
(2) at all other points (i.e., all points except $u_c$ and $u_s$) 
along the boundary of the complex extension of the FM
phase, $M$ vanishes discontinuously.  Result (2) follows because if one starts
in the physical PM phase, a similar analytic continuation argument shows that
$M$ vanishes identically all throughout the complex extension of this phase.  
Inspection of (\ref{msq}) shows that the (real and imaginary parts of the) 
analytic continuation of $M$ are nonzero as one approaches the boundary of 
the complex-extended FM phase from within that phase at points other than 
$u=u_c$ and $u=-1$.  Therefore, $M$ must vanish discontinuously as one 
crosses this boundary from the complex FM to the complex PM phase, as claimed. 

   For the complex-extended AFM phase, we use the well-known symmetry which
holds on loose-packed lattices: under the transformations $K \to -K$ and 
$\sigma_i \to \eta_i \sigma_i$, where $\eta_i=1$ ($-1$) for $\sigma_i$ on the 
even (odd) sublattice, the Hamiltonian is invariant, while the uniform
magnetisation $M$ and the staggered magnetisation $M_{st}$ interchange their 
roles.  The above transformation
takes $k_< \to k_<, \ z \to 1/z, \ u \to 1/u$.  The expression (\ref{msq}) is
invariant and thus describes the staggered magnetisation in the physical AFM 
phase as well as the uniform magnetisation in the physical FM phase. As before,
one generalizes this to a definition of the staggered magnetisation in the
complex-extended AFM phase by analytic continuation from the physical region
$-\infty \le K \le -K_c$ throughout the complex AFM phase, as indicated in
 Fig. 1.  One sees that analogues of the two results which we obtained for 
$M$ can also be derived for $M_{st}$: (1) $M_{st}$ vanishes continuously at two
points on the border of the complex-extended AFM phase, namely,
$u=1/u_c=3+2\sqrt{2}$, the usual PM-AFM critical point, and $u=u_s=-1$, 
the complex-temperature singular
point where $M$ also vanishes; and (2) at all other points on the border of the
complex-extended AFM phase, $M_{st}$ vanishes discontinuously as one crosses
this border into the PM phase.  Note that, as is clear from Fig. 1, the only
point where the complex FM and AFM phases are contiguous is the single point
$u=u_s=-1$, or equivalently, the two points $z = \pm i = -v$. 

   In passing, we note that for both the uniform and staggered magnetisations,
the apparent divergence at $u=1$ plays no role since this point is outside the
two respective regions (complex FM and AFM phases) where the expression
(\ref{msq}) for these quantities applies.

\subsection{The Behaviour of the Inverse Correlation Length as $u \to -1$}
\label{massgap}

\subsubsection{Approach From Within Complex (A)FM Phase}

   For the Ising model on the square lattice, in the physical 
low-temperature phase with real $K$ in the interval $K_c < K \le \infty$, the
asymptotic decay of the row (or equivalently, column) connected correlation 
functions is given by \cite{corr}
\beq
<\sigma_{(0,0)} \sigma_{(0,n)}>_{conn.} \sim n^{-2}e^{-|n|/\xi_{FM,row}}
\label{row}
\eeq
where the 
inverse correlation length (mass gap) is 
\beq
\xi_{FM,row}^{-1} = \ln((v/z)^2) = \ln \biggl [ z^{-2}\biggl ( 
\frac{1-z}{1+z}\biggr )^2 \biggr ] 
\label{xiinv}
\eeq
We now analytically continue this result into the complex extension of the FM
phase and inquire where the mass gap vanishes.  We find that for points within,
and on the border of, the complex-extended FM phase, the mass gap vanishes for
the following set: 
\beq
\xi_{FM,row}^{-1} = 0 \ \ for \ \ z = \{z_c, \ \ \pm i \}
\label{mzero}
\eeq
i.e., the usual PM-FM critical point 
$z_c=\sqrt{2}-1$ and at the two points $z_s = \pm i$ ($u_s=-1$).  
The additional apparent zero at the PM-AFM critical point 
$z=-1/z_c$ is not relevant for the complex FM phase because this point lies 
outside this phase and thus outside the region which can be
reached by analytic continuation of the original formula (\ref{row}); however,
it will be relevant for the inverse correlation length defined within the
complex-extended AFM phase (see below). 
We note the somewhat subtle point that the correct analytic 
continuation of the physical, real-$K$ theory to complex $K$ requires that 
one use $\xi_{FM,row}^{-1}= \ln((v/z)^2)$ as given in eq. (\ref{xiinv}) and 
not $\xi_{FM,row}^{-1} = 2\ln(v/z)$; although these are identical expressions 
for physical $K$, the latter form would miss the zero in $\xi_{FM,row}^{-1}$
at $u=-1$. 

   We now extract the critical exponent(s) for this inverse correlation length
(mass gap) at $z = \pm i$ as these points are approached from within the
complex-extended FM phase.  These exponents (which will turn out to be equal)
are defined by 
\beq
\xi_{FM,row}^{-1} \sim const. \times |z \mp i|^{\nu_{\pm i,row}'}  \ \ for \ \ 
z \to \pm i 
\label{nuidef}
\eeq
from within the FM phase. Expanding $\xi_{FM,row}^{-1}$ about these points 
gives 
\beq
\xi_{FM,row}^{-1} = 0 + 2(\pm i - 1)(z \mp i) + O((z \mp i)^2) 
\label{mexpansion}
\eeq
from which it follows that 
\beq
\nu_{i,row}' = \nu_{-i,row}' = 1
\label{nui'}
\eeq
This motivates the use of a single exponent to describe the singularity at the
single point $u_s=-1$ corresponding to $z = \pm i$, as approached from within
the complex-extended FM phase:
\beq
\nu_{s,row}'= 1
\label{nus'}
\eeq
  
   By the standard argument noted above which shows that on a loose-packed 
lattice such as the square lattice
\beq
\eta_i\eta_j<\sigma_i\sigma_j>(-K)_{conn.} = <\sigma_i\sigma_j>(K)_{conn.}
\label{corfunrel}
\eeq
it follows that the same inverse correlation length (\ref{xiinv}) describes the
asymptotic decay of the connected 2-spin correlation functions along a row or
column in the AFM phase, i.e., formally, 
$\xi_{FM,row}^{-1}=\xi_{AFM,row}^{-1}$, although the same
expression is used in different phases.  Now $\xi_{AFM,row}^{-1}$ vanishes at 
the 
physical PM-AFM critical point $z=-1/z_c$, i.e., $u=1/u_c$, and at the
complex-temperature singular points $z= \pm i$.  It follows that the expansion 
(\ref{mexpansion}) also controls the critical exponent as one approaches the 
points $u=-1$ from the AFM phase (i.e. from the left in Fig. 1(b)).  

   The value of $\nu_{s,row}'$ is evidently the same as the value $\nu'=1$ for
the physical PM-FM critical point, as approached from within the FM phase.
However, one encounters several new features at the complex-temperature
singular point, which we now discuss. 

   One may also extract a correlation length critical exponent from the
asymptotic decay of the diagonal ($d$) correlation function, 
\beq
<\sigma_{(0,0)}\sigma_{(n,n)}>_{conn.} \sim n^{-2}e^{-r/\xi_{FM,d}}
\label{diag}
\eeq
where the distance $r = 2^{1/2}|n|$.  In the physical FM phase, 
\beqs
\xi_{FM,d}^{-1} & = & -2^{-1/2}\ln (k_<^2) \\
           & = & -2^{-1/2}\ln \biggl ( \biggl [\frac{4u}{(1-u)^2} \biggr ]^2 
\biggr )
\label{xidfm}
\eeqs
As with the row (column) spin-spin correlation functions, we may analytically
continue eqs. (\ref{diag})-(\ref{xidfm})
 to apply throughout the complex extension of the FM phase.  
Although the detailed form of $\xi_{FM,d}^{-1}$ is different from that of 
$\xi_{FM,row}^{-1}$, they both have the same complex zeroes within the range 
of this analytic continuation, i.e. the complex FM phase and its boundary.  
It may be recalled that near the physical PM-FM critical point, 
\beq
k_<^2 = 1 + 2(4 + 3\sqrt{2})(u-u_c) + O((u-u_c)^2)
\label{klpmfm}
\eeq
so that the exponent which describes the vanishing of the inverse correlation
length characterizing the diagonal correlation function is that same as that
for the row or column correlation functions.  However, the situation is
different at $u=-1$: using the expansion near $u=-1$ 
(see also eq. (\ref{klexpansion}))
\beq
k_<^2 = 1 - \frac{1}{2}(1+u)^2 + O((1+u)^3)
\label{klsqexpansion}
\eeq
it follows that near $u=-1$, 
\beq
\xi_{FM,d}^{-1} = 2^{-3/2}(1+u)^2 + O((1+u)^3)
\label{xidiag}
\eeq
Hence, the correlation length exponent describing the vanishing of the inverse
correlation length for diagonal correlation functions, at $u=u_s=-1$, as 
approached from within the complex FM phase, is
\beq
\nu_{s,diag.}' = 2
\label{nudiag}
\eeq
i.e., twice the value of the mass gap exponent extracted from the row/column
correlation functions.  This situation is unprecedented for critical exponents
at physical critical points. 

\subsubsection{Approach from Complex PM Phase}

    From an analysis of the asymptotic decays of both the row/column and
diagonal correlation functions in the complex extension of the symmetric, PM
phase, we find that the correlation length does {\it not} diverge as one
approaches the points $v=\pm i$ corresponding to the point $u=-1$ from within
this phase.  This finding is very important, since it implies that the
susceptibility is finite at $u=-1$ when this point is approached from within
the complex PM phase.  

    In the physical PM phase, the row (or column) correlation function has 
the asymptotic decay \cite{corr}
\beq
<\sigma_{(0,0)} \sigma_{(0,n)}> \sim |n|^{-1/2}e^{-|n|/\xi_{PM,row}}
\label{pmrow}
\eeq
where
\beq
\xi_{PM,row}^{-1} = \ln(z/v) = \ln \biggl [ v^{-1}\biggl ( 
\frac{1-v}{1+v}\biggr ) \biggr ] 
\label{xiinvpmrow}
\eeq
As before, we may analytically continue this throughout the complex-extended PM
phase.  The mass gap $\xi_{PM,row}^{-1}$ vanishes only at the physical PM-FM
critical point $v_c=\sqrt{2}-1$.  (The apparent zero at $-1/v_c=-(\sqrt{2}+1)$
is not relevant because this point lies outside the complex-extended PM phase
where the above analytic continuation is valid.)  In particular, as one
approaches the points $v = \pm i$ from within the complex PM phase,
$\xi_{PM,row} \to \ln(-1)$, so that 
$<\sigma_{(0,0)} \sigma_{(0,n)}> \sim (-1)^n |n|^{-1/2}$ as $|n| \to \infty$. 

    We find the same result for the diagonal correlation function, which, in
this complex PM phase, has the asymptotic decay 
\beq
<\sigma_{(0,0)}\sigma_{(n,n)}> \sim |n|^{-1/2}e^{-r/\xi_{PM,d}}
\label{pmdiag}
\eeq
where $r = 2^{1/2}|n|$ and 
\beqs
\xi_{PM,d}^{-1} & = & -2^{-1/2}\ln (k_>) \\
           & = & -2^{-1/2}\ln \biggl ( \frac{4v^2}{(1-v^2)^2} \biggr )
\label{xidpm}
\eeqs
Now near $v = \pm i$, 
\beq
k_> = -1 - (v \mp i)^2 + O(v \mp i)^3)
\label{kgexpansion}
\eeq
Hence, 
although $\xi_{PM,d}^{-1}$ vanishes at the physical critical point 
$v_c$, it is finite at the points $v=\pm i$, where $\xi_{PM,d}^{-1} =
2^{-1/2}\ln(-1)$, so that $<\sigma_{(0,0)}\sigma_{(n,n)}> \sim
|n|^{-1/2}(-1)^n$ at these points, just as was true of the row and column
correlation functions.  
 One thus encounters precisely the type of situation that we discussed 
before in Ref. \cite{ms}, where 
$Re(\xi^{-1})=0$, but $Im(\xi^{-1})$ is nonzero. 
Note that the sum $\sum_{n_0}^{\infty} (-1)^{-n}n^{-1/2}$ (where $n_0$ is an
umimportant lower cutoff) is finite.  Of course, although the correlation
length is finite at $v= \pm i$, as approached from within the PM phase, it is
singular at these points since it is unequal to the value obtained as the
points are approached from a different direction in the complex $v$, $z$, or
$u$ planes.  The asymptotic decay of the
general 2-spin correlation function $<\sigma_{(0,0)}\sigma_{(m,n)}>$ has been
calculated (using Toeplitz determinant methods) \cite{corr}; carrying out an
analytic continuation of this result from the physical PM phase into its
complex extension, we again find that the correlation length is finite at 
$v = \pm i$.  Since a divergence in $\bar\chi$ 
on the border of the (complex extension of the) 
PM phase can only arise from a divergence in the sum over 2-spin correlation 
functions contributing to $\bar\chi$, the above results constitute an analytic
demonstration that the susceptibility is finite at the points $v = \pm i$, as
approached from within the complex PM phase.  This is in sharp contrast to the
approach from within the complex FM or AFM phases, where we have shown that
$\bar\chi$ is divergent.  This type of phenomenon is,
again, to our knowledge, unprecedented in the study of singularities in
thermodynamic functions at physical critical points.  

\subsubsection{A Theorem on $\bar\chi$ }

   From our results in the previous two subsections, using the same reasoning
as in Ref. \cite{ms}, we can infer the following theorem:
\newline

Theorem. The susceptibility $\bar\chi$ has at most finite
non-analyticities on the natural boundary curve (circles in $v$ or $z$,
lima\c{c}on in $u$) separating the complex-extended PM, FM, and AFM phases,
apart from the divergent singularities as one approaches the point
$u_c=3-2\sqrt{2}$ from within either the complex PM or FM phase and the 
point $u_s=-1$ from within the complex FM or AFM phase. 

\section{Scaling Relations and Other Critical Exponents at $u_s=-1$}
\label{scaling}

\subsection{$\alpha_s'+2\beta_s+\gamma_s'=2$}
\label{abgsection}

   Using our result (\ref{gammasconc}) for $\gamma_s'$, together with the
exponents $\alpha_s'$ and $\beta_s$ extracted from the known exact expressions
for $C$ and $M$ in eqs. (\ref{alphas}), (\ref{betas}), we find that the complex
analogue of the scaling relation (from the low-temperature side) 
$\alpha'+2\beta+\gamma'=2$ is satisfied:
\beq
\alpha_s'+2\beta_s+\gamma_s'=2
\label{abg}
\eeq
where the subscript $s$ indicates that this refers to the point $u_s=-1$ and
the primes indicate that the approach to this point is from the
complex-extended broken-symmetry phases, FM or AFM.  To be precise, this
relation is satisfied to within the numerical accuracy of our determination 
of $\gamma_s'$ in eq. (\ref{gammas}) and is satisfied exactly if one uses our
inference in eq. (\ref{gammasconc}) of the exact value of $\gamma_s'$.
However,
our results in the previous section, in particular, the demonstration that
$\bar\chi$ is finite at $v = \pm i$ \ ($u=-1$) as approached from within the
complex PM phase, and hence that $\gamma_s < 0$, already shows that
\beq
\alpha_s+2\beta_s+\gamma_s \ne 2
\label{rush2}
\eeq
i.e. the scaling 
relation for the approach from within the PM phase, is {\it not} valid at 
$u_s$.  We do not know of any extension of the arguments for usual exponent
relations to complex temperature, so it should not be considered a surprise
that such relations do not hold at a complex-temperature singularity. 

\subsection{Hyperscaling Relations}

    Since we have shown above that the inverse correlation length is finite at
the points $v=\pm i$ when approached from within the complex PM phase, the
corresponding exponent $\nu_s < 0$.  Hence, the hyperscaling relation 
$d\nu = 2-\alpha$ does not hold at these points, as approached from within the
complex PM phase.  
Concerning the hyperscaling relation for the approach to
$u_s=-1$ from within the complex FM or AFM phases, namely, 
$d\nu'=2-\alpha'$, if one used the correlation length exponent
$\nu_{s,row}=1$ extracted from the row or column correlation functions, 
then this relation would be satisfied.  However, the situation is more
complicated, since, in particular, $\nu_{s,diag}'=2 \ne \nu_{s,row}'$.

\section{Violation of Universality at $u_s=-1$} 
\label{violations}

    Although the scaling and hyperscaling 
relations discussed above were found to be
satisfied, complex-temperature singularities clearly have different properties
from physical, real-temperature critical points.  Among other things, 
quantities which are real for physical $T$ in general become complex for
complex $T$.  Furthermore, various positivity relations, such as the property
that the specific heat $C > 0$ is not true even when $C$ is real. 
One should therefore be cautious concerning the question of
whether a given property associated with a physical critical point will apply
at a complex-temperature singular point.  Indeed, we shall now demonstrate a
violation of universality at the complex-temperature singular point $u_s$. 

   We recall the meaning of universality as applied to statistical mechanical 
models not involving frustration or competing interactions: 
the universality class, as specified by the critical exponents, depends on (i)
the symmetry group $G$ of the Hamiltonian and the related space of the order
parameter; (ii) the dimensionality of the lattice, but (iii) not on the details
of the Hamiltonian, such as additional spin-spin couplings (provided that 
these are invariant under $G$ and do not introduce frustration or competing 
interactions), and (iv) not on the lattice type (again, provided that this does
not cause frustration).  

   We shall now demonstrate, using exact results, that property (iv) is 
violated at the point $u_s=-1$.  In order to do this, we use the 
expressions for the spontaneous magnetisation on the triangular and 
honeycomb lattices.  These can be written in the same general form 
(\ref{msq}) for the square lattice, but with elliptic moduli which are
different functions of $z$:
\beq
M=\Bigl ( 1-(k_{<,\Lambda})^2 \Bigr )^{1/8}
\label{maggen}
\eeq
where instead of the relation (\ref{kl}) for $k_{<,sq}$, one has
\beq
k_{<,t}=\frac{4z^3}{(1+3z^2)^{1/2}(1-z^2)^{3/2}}
\label{kltri}
\eeq
and 
\beq
k_{<,hc}=\frac{4z^{3/2}(1-z+z^2)^{1/2}}{(1-z)^3(1+z)}
\label{klhc}
\eeq
These apply to the physical FM phases for each lattice, i.e. where 
$0 \le k_< < 1$.  
The explicit forms in terms of the usual low-temperature variable $u$ for 
the triangular and honeycomb lattices are thus \cite{potts}
\beq
M_{t} = \biggl (\frac{1+u}{1-u}\biggr )^{3/8}
\biggl(\frac{1-3u}{1+3u}\biggr)^{1/8} 
\label{mtri}
\eeq
and \cite{naya}
\beq
M_{hc} = \frac{(1+z^2)^{3/8}(1-4z+z^2)^{1/8}}{(1-z)^{3/4}(1+z)^{1/4}}
\label{mhc}
\eeq
which apply within the respective FM phases on these lattices and vanish
elsewhere. 
(Recall that since the honeycomb lattice has an odd coordination number, $q=3$,
$M$ and $\bar\chi$ are not invariant under $z \to -z$ as they are for 
lattices of even $q$.)  We first note a similar feature of the spontaneous
magnetisation on all three lattices: $M$ vanishes continuously at the same
generic set of points, namely the respective PM-FM critical points on each
lattice, and the point $u=-1$, or equivalently the two points $z=\pm i$.  
As is well known, the critical exponent $\beta=1/8$ is the same at the 
respective physical PM-FM critical points.  However, this is {\it not} true at
$u=-1$: for the square lattice, the critical exponent was extracted above as
$\beta_{s,sq}=1/4$, but for the other lattices, 
\beq
\beta_{s,t}=\beta_{s,hc}=\frac{3}{8} \ne \beta_{s,sq}
\label{betatrihex}
\eeq 
Given the fact that $M$ has the same form (\ref{maggen}) in terms of the 
(different) elliptic moduli $k_{<,\Lambda}$ for the three lattices, and given 
that $u_s=-1$ maps to $k_{<,\Lambda}=-1$ for each of these lattices, it 
follows that $M \sim |1+k_{<,\Lambda}|^p$ as $k_{<,\Lambda}$ approaches $-1$ 
from with the FM phase for each case, with the same value $p=1/8$. However, 
this is not the same as the usual meaning of universality, since the 
$k_{<,\Lambda}$ differ as functions of $z$ for each of the three lattices. 

   Since early studies of low-temperature series expansions, it 
has been known that different lattice types have different numbers of
complex-temperature singular points (see, e.g., Ref. \cite{dg} and references
therein).  However, to our knowledge, the obvious violation of universality 
noted above has not been explicitly discussed in the literature.  
Indeed, in an early study of complex-temperature (and
complex-activity) properties of the 3D Ising model \cite{ipz}, from analyses of
low-temperature series expansions on the simple cubic (sc), body-centered cubic
(bcc), and face-centered cubic (fcc) lattices, it was found that the 
numerical evidence was consistent with the equality of critical exponents on 
these three lattices.  The critical exponents at the 
complex-temperature singular point $u \simeq -0.285$ in the Ising model on 
the simple cubic lattice were recently determined to higher precision in 
Ref. \cite{ge}.  It would be useful to calculate longer low-temperature
series for $C$, $M$, and $\bar\chi$ on the bcc and fcc lattices to compare with
the higher-accuracy critical exponents obtained in Ref. \cite{ge} for the sc 
lattice.  However, our exact results on $\beta_s$ already show that
universality does not, in general, hold at complex-temperature critical 
points. 

    Some possible insight into this violation may be gained by remembering 
that even at physical critical points, universality does not, in general, hold
when there is frustration.  One of the earliest examples is the (isotropic)
antiferromagnetic Ising model on the triangular lattice, for which there is no
PM-AFM phase transition at finite $K$.  Accordingly, when examining a given
singular point to see if one could expect universality to hold, one of the
first things which one would necessarily check would be the presence or absence
of frustration, which, in turn, would involve checking whether various spin
configurations only partially minimize the internal energy.  But this initial
check cannot be performed in the usual way at a complex-temperature singular
point, since at such a point the Hamiltonian and internal energy are not, in 
general, real numbers. 

\section{Analysis of High-Temperature Series Expansion for the 
Susceptibility}

   In section 4, as a consequence of our study of the complex-temperature
behaviour of correlation lengths, we showed analytically that the
susceptibility is finite at the points $v=\pm i$ (i.e., $u=-1$) as approached
from within the complex extension of the PM phase.  In this section we shall
carry out a study of high-temperature series expansions for the 
susceptibility.  The results confirm our analytic demonstration and give
further information about $\bar\chi$ at these points. To our 
knowledge, this is the first time that a comparison has been made of the 
behaviour at a complex-temperature singularity as approached from both the 
complex-extended FM (AFM) and PM phases.  For technical reasons, the study of
the high-temperature series in the vicinity of $v=\pm i$ turns out to be 
considerably more difficult than was the case with the low-temperature series
in the vicinity of the equivalent single point $u=-1$.  We begin with a simple
dlog Pad\'{e} study, which is adequate to confirm the absence of a divergent 
singularity; we then proceed to a study with differential approximants. 

  Recall that the high-temperature 
series expansion for the susceptibility is given by 
\beq
\bar\chi = 1 + \sum_{n=1}^{\infty} a_n v^n 
\label{chiv}
\eeq
in terms of the usual high-temperature expansion variable.  We have also
transformed the series to one in the elliptic modulus variable $(k_>)^{1/2}$,
\beq
\bar\chi = 1 + \sum_{n=1}^{\infty} a_n'(k_>)^{n/2}
\label{chikgt}
\eeq
via the relation 
\beq
(k_>)^{1/2}=2v/(1-v^2)
\label{kgt}
\eeq
Note the symmetry
\beq
v \to -1/v \ \Rightarrow (k_>)^{1/2} \to (k_>)^{1/2}
\label{kgtsym}
\eeq
The motivation for using the variable $(k_>)^{1/2}$ is 
the same as was discussed in reference to the low-temperature
series, viz., that the exact expressions for the spin-spin correlation 
functions which actually contribute to the susceptibility are polynomials 
in the complete elliptic integrals of modulus $k_>$ in the PM phase 
(multiplied by algebraic functions of $(k_>)^{1/2}$) \cite{kauff,mpw,gs}.  
Under the
mapping from $v$ to $(k_>)^{1/2}$, the boundaries between the phases transform
as follows: the circle $v = -1+2^{1/2}e^{i\theta}$ is mapped to the right-hand
unit semicircle in the $(k_>)^{1/2}$ plane, i.e., 
$(k_>)^{1/2} = e^{i\phi}$ with $-\pi/2 \le \phi \le \pi/2$. Given the symmetry
(\ref{kgtsym}), this is a two-fold covering; in particular, 
the image of the the PM-FM critical point $v_c$ and the point $-1/v_c$ is the 
single point $(k_>)^{1/2}=1$.  Similarly, the circle  
$v = 1+2^{1/2}e^{i\theta}$ is mapped by a two-fold covering 
to the left-hand unit semicircle, $(k_>)^{1/2} = e^{i\phi}$ with 
$(3/2)\pi \ge \phi \ge \pi/2$. The points $-v_c$ (the PM-AFM critical point)
and $1/v_c$, are taken to the point $(k_>)^{1/2}=-1$.  Finally, the 
points $v=\pm i$ which lie on the intersections of the two circles are mapped 
to $(k_>)^{1/2}=\pm i$, respectively. 

 For the square lattice, the $a_n$ have been calculated to the very high order 
$v^{54}$ by Nickel \cite{nickel}. We have performed a dlog Pad\'{e} analysis 
on this series and have found evidence against a divergence in $\bar\chi$ as
$v$ approaches $\pm i$ from within the PM phase.  Since $\bar\chi$ is real for
real $v$, it follows that if the Pad\'{e} approximant for $d\ln(\bar\chi)/dv$
has a pole at $v=v_0$ with residue $R_0$ at some complex point $v_0$, then
it also has a pole at $v=v_0^*$ with residue $R_0^*$.  Writing the singular
part of $\bar\chi$ as $(1-v/v_0)^{-\gamma_0}$ near $v=v_0$ and recalling that
$R_0=-\gamma_0$, it follows that at the two complex-conjugate poles, the real
parts of the exponents are equal, while the imaginary parts (if nonzero) 
are reversed in sign.  Thus, without loss of generality, it suffices to 
consider only the singularity at $v=i$.  

    We find that the Pad\'{e} approximants
to  $d\ln(\bar\chi)/dv$ yield a reasonably stable pole near to $v=i$, with
$Re(\gamma_{v=i}) < 0$. 
However, even for rather high order $[N/N]$ Pad\'{e} entries with 
$15 \le N \le 23$, the pole position is not as close to the singular point 
as one would require for accurate results; typically, 
$|v_{i,ser.}-i| \simeq 0.08$, much larger than the usual level of 
$O(10^{-3})$ (or better) which one would expect for reasonable accuracy. 

   We have therefore studied the 
equivalent series (\ref{chikgt}) in the elliptic modulus variable, 
$(k_>)^{1/2}$. In terms of the latter  
variable, the leading form of the singularities at $(k_>)^{1/2}=\pm i$  is 
given by 
\beq
\bar\chi_{sing} \sim C_{\pm i}|1 \pm i(k_>)^{1/2}|^{-\gamma_{_{\pm i}}/2}
\label{kgform}
\eeq
since the Taylor series expansion of $(k_>)^{1/2} \mp i$ as a function of 
$v$, near the points $v = \pm i$, starts with the quadratic term:
\beq
(k_>)^{1/2} = \pm i \pm \frac{i}{2}(v \mp i)^2 + O((v \mp i)^3)
\label{kgtexpansion}
\eeq
Of course, the Pad\'{e} approximants exhibit the well-known pole at 
$(k_>)^{1/2}=1$ due to the usual PM-FM critical singularity and the sequence of
poles and zeroes starting near to $(k_>)^{1/2}=-1$ and continuing outward along
the negative real axis attributed to the finite $(1+x)\ln|1+x|$ PM-AFM
singularity, where $x=(k_>)^{1/2}$. 
 For reference, the high order dlog Pad\'{e} entries for the pole position 
near the physical PM-FM critical point get 
this accurate to order $O(10^{-7})$.  They also yield extremely precise
determinations of the exponent, $\gamma$; indeed, among the $[N/N]$ entries
with $N$ around 20, the values of $\gamma$ only differ from $1.75$ by amounts
of order $10^{-4}-10^{-5}$.  As regards complex-temperature singularities, 
the approximants exhibit two poles which converge to 
$(k_>)^{1/2}=\pm i$.  These, together with the values of $Re(\gamma_i)$, are
shown in Table 8. 
\begin{table}
\begin{center}
\begin{tabular}{|c|c|c|c|} \hline \hline  & & & \\
$[N/N]$ & $(k_>)^{1/2}_{i}$ & $|(k_>)^{1/2}_i-i|$ & $Re(\gamma_i)$ \\
 & & & \\
\hline \hline
[7/7]   & $0.972846i+0.032631$  & $4.2 \times 10^{-2}$ & $-0.4297$ 
\\ \hline
[8/8]   & $1.00343i+0.034469$   & $3.5 \times 10^{-2}$ & $-0.5749$ 
\\ \hline
[9/9]   & $1.00632i+0.015344$   & $1.7 \times 10^{-2}$ & $-0.7720$  
\\ \hline
[10/10] & $1.00729i+0.003183$   & $0.80 \times 10^{-2}$  & $-0.83235$  
\\ \hline
[11/11] & $0.997106i+0.001684$  & $3.3 \times 10^{-3}$  & $-0.6024$  
\\ \hline
[12/12] & $0.999979i+0.000388$ & $3.9 \times 10^{-4}$  & $-0.6432$  
\\  \hline
[13/13] & $0.999715i+0.000650^*$ & $7.1 \times 10^{-4}$ & $-0.6404^*$  
\\ \hline
[14/14] & $1.00053i+0.006735$   & $6.8 \times 10^{-3}$ & $-0.7166$  
\\  \hline
[15/15] & $1.00026i+0.006752^*$   & $6.8 \times 10^{-3}$ & $-0.7101^*$
\\ \hline 
[16/16] & $0.998362i+0.003564$   & $3.9 \times 10^{-3}$ &  $-0.6380$
\\ \hline
[17/17] & $0.998506i+0.003520^*$  & $3.8 \times 10^{-3}$ & $-0.6409^*$
\\ \hline
[18/18] & $0.999870i+0.003009$  & $3.0 \times 10^{-3}$ & $-0.6727$
\\ \hline
[19/19] & $0.998932i+0.002885$  & $3.1 \times 10^{-3}$ & $-0.6442$
\\ \hline
[20/20] & $0.999033i+0.002739^*$  & $2.9 \times 10^{-3}$ & $-0.6449^*$
\\ \hline
[21/21] & $0.999021i+0.002416$  & $2.6 \times 10^{-3}$ & $-0.6391$
\\ \hline
[22/22] & $0.999472i+0.002638$  & $2.7 \times 10^{-3}$ & $-0.6567$
\\ \hline
[23/23] & $0.999487i+0.002646$  & $2.7 \times 10^{-3}$ & $-0.6573$
\\ \hline
[24/24] & $0.999162i+0.002051$ & $2.2 \times 10^{-3}$ & $-0.6374$
\\ 
\hline \hline
\end{tabular}
\end{center}
\caption{Values of $(k_>)^{1/2}_i$ and $Re(\gamma_i)$ from diagonal dlog 
Pad\'{e} approximants to high-temperature series for $\bar\chi$ starting 
with the series to $O((k_>^{1/2})^{15})=O(v^{15})$.}
\label{table8}
\end{table}
The values of $Re(\gamma_i)$ are stable and are negative, indicating that 
$\bar\chi$ has finite singularities at $v=\pm i$ as these points are 
approached from within the complex PM phase.  (The values of $Im(\gamma_i)$ 
will be discussed below.)  This confirms the conclusion of our analytic study
in section 4, viz., that $\bar\chi$ is finite at $v=\pm i$ as approached from
within the complex PM phase.  (That it is singular is obvious since it has
different values when approached from different directions in the complex $v$,
$z$, or $u$ plane.) 

     To check the sensitivity of the dlog Pad\'{e} analysis
to possible divergent singularities, we have studied the test function 
\beq
\bar\chi(x)_{test} = A_{c,x}(1-x)^{-7/4} - A_{AFM,x}(1+x)\ln(1+x) + 
A_{s,x}(1+x^2)^{-\gamma_s/2} + B
\label{ftest}
\eeq
where $x=(k_>)^{1/2}$.  This function incorporates the known PM-FM and 
PM-AFM singularities, a hypothetical divergent singularity at 
$(k_>)^{1/2}=v=\pm i$ with exponent corresponding to $(1+v^2)^{-\gamma_s}$, 
and an additional background term $B$.  We have used the known critical 
amplitudes 
$A_{c,x}=2^{7/8}A_{c,v} = 1.4153665...$, where $A_{c,v}$ is defined by 
$\bar\chi_{sing} \sim A_{c,v}(1-v/v_c)^{-7/4}$ for $v \nearrow v_c$ and is 
given by $A_{c,v}=0.77173406...$ \cite{conflu1,kc2}; and 
$A_{AFM,x}=2^{1/2}A_{v,AFM}=0.28$, where $A_{v,AFM}$ is
defined by $\bar\chi_{sing} \sim -A_{v,AFM}(1+v/v_c)\ln|1+v/v_c|$ for $-v
\nearrow -v_c$ and is given by $A_{v,AFM} \simeq 0.20$ \cite{kc2}.  We have 
varied $A_{s,x}$ and $B$ over a range of values and $\gamma_s$ over the range
$1/4 < \gamma_s \le 3/2$, and have found 
that if these quantities have values such that the dlog Pad\'{e} 
approximants locate the singularities at $v = (k_>)^{1/2}= \pm i$ with an 
accuracy comparable to that which we observe in Table 8, then the approximants 
also yield reasonably accurate values for $\gamma_s$. 
In particular, if we make the values of $A_{s,x}$, $\gamma_s$ and/or $B$ so 
small that the Pad\'{e} 
fails to yield an accurate value for $\gamma_s$, then it also fails to locate
the singularities at $(k_>)^{1/2}=\pm i$ in the test function with the 
accuracy that it does successfully locate them for the actual $\bar\chi$.  
Hence, the dlog Pad\'{e} would not miss divergent singularities at
$v = (k_>)^{1/2}=\pm i$ if they were really present in $\bar\chi$. 

However, the dlog Pad\'{e} method is not, in general, satisfactory for 
a finite singularity, and hence, given that $Re(\gamma_i) < 0$, 
one cannot, {\it a priori}, trust the actual values of $\gamma_i$ which it
yields. The appropriate technique to investigate such finite singularities in
the presence of background terms is provided by differential approximants (DA)
\cite{tonyg,gj,early,hb,mef,rjg}.  To be precise, if a function of a generic 
variable $\zeta$ is of the product form 
$f(\zeta) = \prod_{j=1}^N(1-\zeta/\zeta_j)^{-p_j}$, then the dlog Pad\'{e}
method is, in principle, adequate to obtain the positions and exponents for the
singularities even if some $p_j < 0$.  If $f$ has an additive 
background term near a singularity, i.e., if $f_{sing} \sim
A(1-\zeta/\zeta_j)^{-p_j} + B(\zeta)$ with $B(\zeta)$ analytic near $\zeta_j$,
then, if a given $p_j > 0$, the first Darboux theorem \cite{pade} shows that
for sufficiently high orders, this will dominate over $B(\zeta)$ so that the 
dlog Pad\'{e} method can still be satisfactory, but if the given $p_j < 0$, 
then one should use differential approximants. 

    We recall that in the differential approximant method the function 
$f = \bar\chi$ being
approximated satisfies a linear ordinary differential equation (ODE) of 
$K'th$ order, ${\cal L}_{{\bf M},L}f_K(\zeta) =
\sum_{j=0}^{K}Q_{j}(\zeta)D^j f_K(\zeta) = R(\zeta)$, where 
$Q_j(\zeta) = \sum_{\ell=0}^{M_j} Q_{j,\ell}z^{\ell}$ and 
$R(\zeta)=\sum_{\ell=0}^L R_\ell \zeta^\ell$ \cite{tonyg,hb,mef,rjg}. In one 
implementation of the method \cite{hb,mef}, $D \equiv d/d\zeta$, while in
another \cite{rjg,tonyg}, $D \equiv \zeta d/d\zeta$.  We adopt the choice for 
$D$ used in Ref. \cite{rjg,tonyg}; these authors have found that both 
choices give comparable results.  The solution to 
this ODE, with the initial condition $f(0)=1$, is 
the resultant approximant, labelled as $[L/M_0;...;M_K]$.  The general solution
of the ODE has the form $f_j(\zeta) \sim A_j(\zeta)|\zeta-\zeta_j|^{-p_j} +
B(\zeta)$ for $\zeta \to \zeta_j$.  The singular points $\zeta_j$ are
determined as the zeroes of $Q_K(\zeta)$ and are regular singular points of
the ODE, and the exponents are given by $-p_j = K-1-Q_{K-1}(\zeta_j)/(\zeta_j
Q_K'(\zeta_j))$. Further details on the method can be found in Refs. 
\cite{tonyg,hb,mef,rjg}.  For an extrapolation procedure to be discussed below,
we shall use a number of poles at different positions close to the 
singularity; for this reason, we use unbiased differential approximants.  
Studying the susceptibility series in the variable $v$, we find that 
the differential approximants do not yield singularities sufficiently
close to $\pm i$ to be accurate, just as was true of the dlog Pad\'{e}
method (which is a special case of DA).  As before, we have obtained
considerably better results with the series in the elliptic modulus variable 
$(k_>)^{1/2}$.  We have calculated the $K=1$ differential approximants 
$[L/M_0;M_1]$ for $4 \le L \le 24$ and $10 \le M_0 \le 20$ with $M_1=M_0, \ M_0
\pm 1$ subject to the constraint $L+M_0+M_1+1 \le 49$ (terms up to $O(v^{49}) =
O((k^{1/2}_>)^{49})$ were used).  Many of the poles may reflect finite
singularities along the arcs of the circles bounding the complex PM phase, as
discussed previously \cite{ms}.  To consider a pole to represent the
singularity at $(k_>)^{1/2}=i$, we require that its distance from this point
satisfy $|(k_>)^{1/2}-i| < 1 \times 10^{-2}$.  Secondly, we require that the
pole position lie within the circle $|k_>| = 1$ which forms the boundary of the
complex PM phase, since the high-temperature series only converges to
$\bar\chi$ within this boundary.  In Table 9 we display the 
$K=1$ differential approximants with $L$ even which satisfy these conditions. 
(The approximants with odd $L$ are not listed to save space; they
yield conclusions in agreement with those obtained from the approximants with
even $L$.) 
\begin{table}
\begin{center}
\begin{tabular}{|c|c|c|c|} \hline \hline  & & & \\
$[L/M_0;M_1]$ & $(k_>)^{1/2}_{i}$ & $|(k_>)^{1/2}_i-i|$ & $\gamma_i$ \\
 & & & \\
\hline \hline
[4/18;19] & $0.993373i+0.004293$ & $0.79 \times 10^{-2}$ &  $-1.420-1.079i$ 
\\ \hline
[4/18;20] & $0.991349i+0.002545$ & $0.90 \times 10^{-2}$ & $-1.820-0.8796i$ 
\\ \hline 
[8/18;16] & $0.990874i+0.003398$ & $0.97 \times 10^{-2}$ & $-1.950-0.3293i$ 
\\ \hline
[8/18;17] & $0.988550i-0.006664$ & $1.3 \times 10^{-2}$ & $-1.944+1.741i$ 
\\ \hline
[8/18;18] & $0.990839i+0.004293$ & $1.0 \times 10^{-2}$ & $-2.164-0.3726i$ 
\\ \hline
[10/16;18] & $0.999122i-0.001421$ & $1.7 \times 10^{-3}$  & $-0.8568+0.4512i$ 
\\ \hline 
[12/10;10] & $0.9987905i-0.005881$ & $0.60 \times 10^{-2}$ & $-1.243+0.4854i$ 
\\ \hline
[14/10;10] & $0.990038i+0.007285$ & $1.2 \times 10^{-2}$ & $-2.294-1.181i$ 
\\ \hline
[14/12;14] & $0.990566i-0.001845$ & $0.96 \times 10^{-2}$ & $-1.611+0.3934i$ 
\\ \hline
[18/10;10] & $0.990270i+0.002944$ & $1.0 \times 10^{-2}$ & $-1.907-0.3510i$ 
\\ \hline 
[18/12;12] & $0.998597i+0.01127$ & $1.1 \times 10^{-2}$  & $-0.7024-1.417i$ 
\\ \hline 
[20/10;11] & $0.994749i-0.001108$ & $0.54 \times 10^{-2}$ & $-1.267+0.3876i$ 
\\ \hline 
[20/10;12] & $0.995176i-0.002265$ & $0.53 \times 10^{-2}$ & $-1.214+0.5585i$ 
\\ \hline 
[20/12;10] & $0.998042i+0.008232$ & $0.85 \times 10^{-2}$ & $-0.1837-1.433i$ 
\\ \hline 
[20/12;11] & $0.995593i-0.003949$ & $0.59 \times 10^{-2}$ & $-1.105+0.9183i$ 
\\ \hline
[20/12;12] & $0.998991i-0.000353$ & $1.1 \times 10^{-3}$ & $-0.6833+0.3278i$ 
\\ \hline
[22/10;11] & $0.997787i-0.002213$ & $3.1 \times 10^{-3}$ & $-0.8149+0.6317i$ 
\\ \hline
[22/10;12] & $0.998212i-0.001353$ & $2.2 \times 10^{-3}$ & $-0.7283+0.5223i$ 
\\ \hline 
[24/10;9]  & $0.993917i-0.006415$ & $0.88 \times 10^{-2}$ & $-1.850+0.9510i$ 
\\ \hline 
[24/10;10] & $0.997266i-0.001653$ & $3.2 \times 10^{-3}$ & $-0.9898+0.4162i$ 
\\ \hline 
[24/10;11] & $0.997902i+0.0004882$ & $2.2 \times 10^{-3}$ & $-0.5841-0.0597i$ 
\\ \hline 
[24/10;12] & $0.997133i-0.008122$ & $0.86 \times 10^{-2}$ & $-1.120+1.803i$ \\
\hline \hline
\end{tabular}
\end{center}
\caption{Values of $(k_>)^{1/2}_i$ and $\gamma_i$ from $K=1$ differential
approximants to high-temperature series for $\bar\chi$.  See text for
definition of $[L/M_0,M_1]$ approximant.}
\label{table9}
\end{table}
Evidently, there is a large scatter of values of $\gamma_i$.  Also, the
distances from the singularity are usually larger than the $O(10^{-3})$ level
which one would normally consider necessary for accurate results.  However, we
can still draw useful information from this table.  First, all of the values 
of $\gamma_i$ satisfying the two conditions above have the property 
that $Re(\gamma_i) < 0$.  Second, the approximants in Table 9 which yield 
poles closest to $(k_>)^{1/2}=i$, namely $[10/16;18]$, $[20/12;12]$, 
$[22/10;12]$, and $[24/10;11]$, do give roughly consistent values of 
$Re(\gamma_i)$.  If we plot the values of
$Re(\gamma_i)$ as a function of the distance $|(k_>)^{1/2}-i|$ and extrapolate 
to zero distance from the singularity, we obtain $Re(\gamma_i) \simeq -0.65$.
This is consistent with the values obtained for this quantity from the dlog 
Pad\'{e} study. 

    We can make further progress by noticing an important correlation: the 
sign of $Im(\gamma_i)$ is opposite to the sign of
the deviation $Re((k_>)^{1/2}_i)$ from zero.  Indeed, when we plot the 
values of $Im(\gamma_i)$ as a function of $Re((k_>)_i)$, they can be roughly 
fit to a line going through zero when $Re((k_>)^{1/2}_i)=0$.  But the
singularity which we are studying is at $(k_>)^{1/2}=i$, so we are led to the 
tentative conclusion that for this singularity, $Im(\gamma_i)=0$. 
It follows also that for the conjugate singularity at $(k_>)^{1/2}=-i$, 
$Im(\gamma_{-i})=0$, whence $\gamma_i = \gamma_{-i} = \gamma_s$.  As regards
the values of $Im(\gamma_i)$ from the dlog Pad\'{e} study, one can see that 
all of the high-order diagonal approximants are characterized by the same
(positive) sign for $Re((k_>)^{1/2}$, i.e. the pole positions are slightly 
to the right of $(k_>)^{1/2}=i$.  This correlates with the observed feature 
that these Pad\'{e} approximants have the stable nonzero negative values 
$Im(\gamma_i) \simeq -0.25$.  
Indeed, when we examine other poles in the dlog Pad\'{e} approximants
reasonably close to the point $(k_>)^{1/2}=i$, we observe the same correlation
that $Im(\gamma_i)$ has a sign opposite to that of $Re((k_>)^{1/2})$. 
This suggests that if we had a reasonably large set of such nearby poles, 
then we could carry out an extrapolation similar to the one that we performed 
with the differential approximants.  There are not enough close poles with
$Re((k_>)^{1/2}) < 0$ to do 
this accurately, but one can say that such an extrapolation is crudely
consistent with $Im(\gamma_i)=0$. 

   Summarizing, then, the results of our analysis with differential 
approximants, like those of the simple dlog Pad\'{e} study, confirm our
analytic demonstration that $\bar\chi$ is finite (although, of course,
singular) at the points 
$v=(k_>)^{1/2}= \pm i$ (i.e., $u=-1$) when approached from within the complex 
PM phase. Furthermore, if we restrict to the four poles closest to 
$v=(k_>)^{1/2}= i$, the DA's yield values for $Re(\gamma_i)$ which are roughly
mutually consistent.  Note that for these four poles,
the distance from the singularity, $2-3 \times 10^{-3}$, is comparable to 
that for the poles produced by the high-order dlog Pad\'{e} approximants in
Table 8.  Moreover, when we extrapolate the values of $Re(\gamma_i)$ to apply
precisely at the location of the singularity, we obtain a value consistent with
that from the dlog Pad\'{e} approximants.  Using a similar extrapolation
procedure with the differential approximants, we infer that $Im(\gamma_i)=0$.

    We recall that in principle, $\bar\chi$ may have singularities elsewhere on
the arcs forming the natural boundary of the complex PM phase, as discussed in
Ref. \cite{ms}.  It was proved there that if such singularities exist, they
must be finite.  Indeed, from a simple dlog Pad\'{e} study, it was observed 
in that paper that the poles lay close to these arcs.  The differential 
approximants presumably give some information about these possible finite
singularities along the arcs.  We shall leave the more detailed investigation 
of the behaviour on these arcs to future work.  However, from our present
results, we may make some interesting observations. To the extent that the 
poles in the differential approximants away from $(k_>)^{1/2}=\pm i$ lie near
to the arcs bounding the complex PM region, they might reflect finite
singularities in $\bar\chi$ along these arcs.  The results of our study then
suggest that the associated exponents would have nonzero imaginary parts, at
points along the arcs in $v$ or circle in $(k_>)^{1/2}$ apart from the points
$(k_>)^{1/2}=\pm 1$ ($v= \pm v_c$) and $(k_>)^{1/2}=\pm i$ ($v= \pm i$). 
In assessing this possibility, it is, of course, incumbent upon one first 
to check whether there are any rigorous
theorems forbidding this.  We have not been able to derive any such theorem. 
Of course, the usual rigorous theorems governing the behaviour of 
thermodynamic quantities and, in particular, their critical exponents, 
assume physical values of the temperature. Two properties which must be
satisfied are that $\bar\chi$ must be real and positive for physical
temperature.  These two properties do not exclude the existence of a complex 
exponent at a complex-temperature singularity in $\bar\chi$.  
To demonstrate this, we
consider a generic form for the singular part of $\bar\chi$ with such
singularities: 
\beq
\bar\chi_{sing}(\zeta) = A_j (1-\zeta/\zeta_0)^{-\gamma_0} + 
 A_0^* (1-\zeta/{\zeta_0^*})^{-\gamma_0^*}
\label{chiexample}
\eeq
where $\zeta$ denotes a generic variable ($v$, $(k_>)^{1/2}$ $z$, etc.),
$Im(\gamma_0) \ne 0$,  and the
form (\ref{chiexample}) applies, say, for $|\zeta| < |\zeta_0|$. 
Evidently, this has the property that 
\beq
\bar\chi_{sing}(\zeta^*) = \bar\chi_{sing}(\zeta)^*
\label{reality}
\eeq
which implies that if $\zeta$ is real, then $\bar\chi(\zeta)$ is also 
real.\footnote{
Actually, real $\zeta$ may still correspond to complex temperature; for
example, as eq. (\ref{shift1}) shows, 
$K = r + n i \pi$ with $r$ real corresponds
to real $u$ and hence real $\bar\chi$.  The symmetries (\ref{shift1}) and
(\ref{shift2}) imply that for the indicated infinite sets of complex values of
$K$, quantities such as $\bar\chi$ are still real.} 
We have also checked that one can write down examples of
$\bar\chi_{sing}$ of the form in eq. (\ref{chiexample}) which yield not just
real, but positive, $\bar\chi$ for physical temperature. 
If $\bar\chi$ has a singularity at the point $\zeta = \zeta_0$ of the leading 
form in eq. (\ref{chiexample}) 
where the real and imaginary parts of $\gamma_0 $ are given by 
\beq
\gamma_0 = \gamma_{_R} + i \gamma_{_I}
\label{gammacartesian}
\eeq
then for $\zeta = \zeta_0(1+\epsilon e^{i\theta})$, as $\epsilon \to 0$,
\beq
\bar\chi \sim |\epsilon|^{-\gamma_{_R}} \biggl [ \cos(\gamma_{_I} 
\ln(|\epsilon|)) - i \sin(\gamma_{_I} \ln(|\epsilon|)) \biggr ]
\label{fullchivsing}
\eeq
Given our theorem above that 
that $\bar\chi$ must be finite on the arcs away from $v=v_c$ and $v=\pm i$
(approached from within the FM or AFM phases), 
it follows that $\gamma_{_R}$ must be negative away from these
points.  The effect of the nonzero $\gamma_{_I}$ is to produce an infinitely 
rapid oscillation in $\bar\chi$ as
$\zeta \to \zeta_0$ with $|\zeta| < |\zeta_0|$.  
Although the points on the arcs are not isolated
singularities, but instead, points on the natural boundaries between the
complex-extended PM, FM, and AFM phases, this behaviour is quite analogous to
the infinitely rapid oscillations at isolated essential singularities such as
in the functions $\zeta \sin(1/\zeta)$ or $\zeta \cos(1/\zeta)$ at $\zeta=0$.
Given our results that $\bar\chi$ diverges at $u=-1$ as one approaches this
point from the complex FM or AFM phases, but has a finite singularity 
as one approaches the
point from the complex PM phase, one may observe that 
this behaviour is somewhat reminiscent of the function
$\exp(1/\zeta)$, which has an (isolated) essential singularity at $\zeta=0$ and
diverges (vanishes) when this point is approached from the positive (negative)
real axis. 

    A physical argument for the divergence in the susceptibility at a
second-order phase transition, such as the PM-FM critical point, is to note 
that the magnetisation vanishes continuously as one approaches this 
point from within the FM phase, so that, in the limit, 
an arbitrarily small external field $H$ has an arbitrarily large
relative effect on the resultant magnetisation $M(H)$.  This argument also
agrees with the divergence in the susceptibility that we found at the 
complex-temperature point $u=-1$, as approached from within 
the complex-extended
 FM phase, since this is the only other point at which $M$ vanishes
continuously, starting from within the complex FM phase.  For the physical
PM-FM critical point the same argument motivates the divergence in $\bar\chi$
as one approaches this point from within the PM phase. However, our analytic
demonstration (and confirmation via high-temperature series analysis)
that $\bar\chi$ has a finite singularity when one 
approaches $u=-1$ from within the complex PM phase shows that for
complex-temperature singularities, this reasoning for physical temperatures
is not applicable in a naive manner.

\section{Conclusions}

   In summary, we have carried out a study of the complex-temperature
singularities of the susceptibility of the 2D Ising model on a square 
lattice.  From an analysis of low-temperature series expansions, we have found 
evidence that as one approaches the point $u=u_s=-1$ from within the complex 
extensions of the physical ferromagnetic or antiferromagnetic phases, 
$\bar\chi$ has a divergent singularity. Our results are consistent with the 
conclusion that the critical exponent for this singularity is 
$\gamma_s'=3/2$.  We have also calculated the critical amplitude.  
However, from an analysis of the asymptotic decays of spin-spin correlation 
functions, we have shown that the correlation length is finite, and hence the
susceptibility is finite (although both are singular) at the points
$v=\pm i$ corresponding to $u=-1$ when 
approached from within the complex-extended symmetric, paramagnetic phase. 
This is confirmed by a study of high-temperature series expansions. 

     Our results are of interest because they 
elucidate the behaviour of the susceptibility as an analytic function. 
The goal of calculating the susceptibility of the 2D Ising model 
(or even making a conjecture for this function which agrees with available 
series expansions) has remained elusive for half a century since Onsager's 
calculation of the free energy.  Our results should be useful for this quest 
because they provide new properties which must be satisfied by such a
conjecture or calculation. 

   One can think of a number of further related studies to perform.  In one
direction, 
using low- and high-temperature expansions, we have carried out analyses with 
dlog Pad\'{e} and differential approximants to 
investigate complex-temperature singularities of $\bar\chi$ in the Ising
model on the triangular and honeycomb lattices.  Our results will be reported
in a sequel paper.   One could also examine other models like the $d=2$, 
$q=3$ Potts model. Yet another topic is how ideas of conformal
field theory, which have greatly illuminated the critical behaviour at the
physical critical points of various 2D models, can be applied to this
complex-temperature singularity.  Clearly, there is much work for the future.

   This research was supported in part by the NSF grant PHY-93-09888.

\vfill

\eject

\begin{center}
{\bf Figure Caption}
\end{center}

\vspace{5mm}

Fig. 1. \  Phases and associated boundaries in the complex variables (a) $v$, 
(b) $z$, and (c) $u$, as defined in eqs. (\ref{v})-(\ref{u}).  In the variable
$k_<$ defined in (\ref{kl}), the complex FM and AFM phases are mapped into the
interior of the unit circle, and the PM phase to the exterior of this circle. 

\vfill

\eject

\end{document}